\documentclass[]{aastex631}

\submitjournal{ApJS}

\usepackage{graphicx} 
\usepackage{xcolor}
\usepackage{color}                       

\usepackage[utf8x]{inputenc}
\usepackage{booktabs}

\usepackage{soul,color}

\begin{document}

\title{Catalogue of Solar Radio Bursts Detected by the LOFAR LV614 Station at the Irbene Observatory during Solar Cycle 25}

\correspondingauthor{Jānis Šteinbergs}
\email{janis.steinbergs@venta.lv}

%


\author[0000-0001-9265-3655]{Jānis Šteinbergs}
\affiliation{Riga Technical University, 6A Kipsalas Street, Riga, LV-1048, Latvia}
\affiliation{Ventspils University of Applied Sciences, Engineering Research Institute \lq\lq Ventspils International Radio Astronomy Center\rq\rq, Ventspils, Latvia}

\author[0000-0002-0687-6172]{Dmitrii Kolotkov}
\affiliation{Centre for Fusion, Space and Astrophysics, Department of Physics, University of Warwick, Coventry, CV4 7AL, UK}
\affiliation{Ventspils University of Applied Sciences, Engineering Research Institute \lq\lq Ventspils International Radio Astronomy Center\rq\rq, Ventspils, Latvia}
\email{D.Kolotkov.1@warwick.ac.uk}

\author[0000-0001-6423-8286]{Valery M. Nakariakov}
\affiliation{Centre for Fusion, Space and Astrophysics, Department of Physics, University of Warwick, Coventry, CV4 7AL, UK}
\affiliation{Centro de Investigacion en Astronom\'ia, Universidad Bernardo O'Higgins, Avenida Viel 1497, Santiago, Chile}
\affiliation{Ventspils University of Applied Sciences, Engineering Research Institute \lq\lq Ventspils International Radio Astronomy Center\rq\rq, Ventspils, Latvia}
\email{V.Nakariakov@warwick.ac.uk}

\author[0000-0003-3655-2280]{Vladislavs Bezrukovs}
\affiliation{Ventspils University of Applied Sciences, Engineering Research Institute \lq\lq Ventspils International Radio Astronomy Center\rq\rq, Ventspils, Latvia}
\email{vladislavsb@venta.lv}

\author[0000-0002-6760-797X]{Pietro Zucca}
\affiliation{ASTRON – Netherlands Institute for Radio Astronomy, Oude Hoogeveensedijk 4, 7991 PD Dwingeloo, The Netherlands}
\email{zucca@astron.nl}

\author[0000-0001-6345-8084]{Jānis Kaminskis}
\affiliation{Riga Technical University, 6A Kipsalas Street, Riga, LV-1048, Latvia}
\email{janis.kaminskis@rtu.lv}

\begin{abstract}

A catalogue\footnote{the DOI for the catalogue \url{https://doi.org/10.5281/zenodo.20157853}} of solar radio bursts detected with the LV614 LOFAR station at the Irbene Observatory, Latvia, in the stand-alone mode from June 2022 to August 2025 in the 29--60~MHz frequency range is presented.
Observations were conducted with the weekly cadence, one-second time resolution, and 0.195 MHz
spectral resolution.
The cumulative duration of the observation was 1125.2 h.
For all analysed observations, bandpass calibration was applied using Cassiopeia A as a
reference source.
In total, the catalogue includes 335 solar radio bursts.
Radio bursts of all five main types, I--V, were detected. Specifically, the catalogue contains parameters of 23, 8, 293, 6, and 5 bursts of Type I, II, III, IV, and V, respectively.
The detected burst peak flux ranges from 0.3 SFU to 234 SFU.
The catalogue provides information about the peak flux and frequency, and the estimated height of the source above the solar surface. For Type II and III bursts, we also give the estimations of the  frequency drifts, with mean values $ 0.043 ~\mathrm{MHz\,s^{-1}}$ and $ 4.383 ~\mathrm{MHz\,s^{-1}}$, respectively. {We also estimated the fit $df/dt = Af^a$, with the best fit parameters $A=0.78$ and $a=0.43$.} The numbers of Type I and III events are found to decrease with the peak flux, with the power law indices about $-0.9$ and $-1.25$, respectively. {For Type III, we estimated electron beam speed ranging from 0.005 to 0.477 of the speed of light for the fundamental frequency, and 0.008 to 0.794 of the speed of light for its second harmonic.
}  

\end{abstract}

\keywords{Solar activity (1475) --- Solar flares (1496) --- Solar coronal mass ejections (310) --- Solar radio emission (1522)}

\section{Introduction} \label{Introduction}

Solar activity plays a central role in driving space weather. Among other impulsive energy releases on the Sun, solar radio bursts can directly affect navigation systems and telecommunications \citep{giersch2017solar}. Furthermore, the most powerful solar bursts can cause significant disturbances in the upper atmosphere, as reported by \citet{ghidoni2025lofar}.
In addition, monitoring solar radio bursts across a range of frequencies provides important information for the development of space weather forecasting models, particularly for the prediction of geomagnetic storms \citep{akhavan2025sunrise}, which in turn enables the forecasting of ionospheric disturbances.

Solar radio bursts are a ubiquitous manifestation of solar activity, commonly observed during eruptive events such as solar flares and coronal mass ejections (CMEs) \citep[e.g.,][]{morosan2022exploring}. Solar radio bursts are traditionally classified into five main types based on their characteristic signatures in dynamic spectra \citep{liu2022interferometric}. Type I bursts appear as short-durational, narrowband radio enhancements associated with active regions \citep{melrose1975plasma}. Type II bursts exhibit slow frequency drifts and are generally attributed to electrons accelerated at shock fronts propagating through the corona, often in association with CMEs \citep{2023A&A...675A.102K}. In contrast, Type III bursts are characterised by rapid frequency drifts, reflecting the propagation of electron beams along open magnetic field lines \citep{2021SoPh..296...57K}. Type IV bursts form broadband continuum emissions with complex and highly variable temporal structure \citep{morosan2019variable}, while Type V bursts are weaker continuum emissions that typically follow Type III activity \citep{morosan2014lofar}. 

The LOw-Frequency ARray (LOFAR) \citep{van2013lofar} is particularly well suited for observations of solar radio bursts due to its coverage of the low-frequency range (10--240~MHz), which corresponds to plasma emission originating in the solar corona and inner heliosphere where such bursts are generated.  The high time and frequency resolution of LOFAR enables detailed studies of the rapid temporal variability and fine spectral structure characteristic of solar radio emission. In addition, LOFAR may provide imaging spectroscopy capability, allowing the simultaneous investigation of the spatial, temporal, and spectral evolution of radio sources, which is important for tracing the propagation of energetic electrons and shock fronts associated with solar eruptions. The multi-beam design further enables simultaneous observations of the Sun and calibration sources, facilitating reliable flux measurements. These capabilities make LOFAR a powerful instrument for advancing the understanding of the physical mechanisms underlying solar radio bursts.

Observations with LOFAR have significantly advanced the study of solar radio bursts at low frequencies. High time–frequency resolution measurements have enabled detailed investigations of Type III bursts, revealing fine spectral structures such as striae and facilitating the tracing of electron beam propagation through the corona and interplanetary medium \citep[e.g.,][]{2015A&A...580A..65M, 2017A&A...606A.141R, 2017NatCo...8.1515K, 2018ApJ...861...33K, 2018SoPh..293..115S}. LOFAR imaging spectroscopy has also provided new insights into Type II bursts, including the localisation and evolution of shock-associated radio sources and their relationship to coronal mass ejections \citep[e.g.,][]{2018A&A...615A..89Z, 2019NatAs...3..452M, 2020ApJ...897L..15M, 2025A&A...695A..70M,2025A&A...703A.271Z}. 
In addition, observations of Type IV continua have uncovered complex and rapidly evolving source morphologies, highlighting the role of magnetic structures and energetic electrons in their generation \citep[e.g.,][]{2022SoPh..297..115L}. 

The aim of this paper is to present a catalogue of solar radio bursts detected with the LOFAR LV614 station. This catalogue presents the first scientific study of solar radio bursts conducted at the Irbene radio telescope complex. 

Since the commissioning of the LOFAR LV614 station in September 2019, the observatory has gained the capability to investigate solar radio bursts.  Observations in single-station mode enable the study of solar radio bursts through their signatures in dynamic spectra. Absolute flux calibration and, hence, quantitative estimates of burst intensities are achieved by simultaneous observations of the Sun and the supernova remnant Cassiopeia A, which serves as a flux calibrator with well-established spectral properties \citep{perley2017accurate}. In Section~\ref{Sec:obs}, we describe the observations and data processing methodology. 
Section~\ref{Sec:stat} presents statistical properties of the detected radio bursts. Conclusions are given in Section~\ref{Sec:con}. 
The catalogue of detected radio bursts with parameters is presented in Appendix.



\section{Observations and data processing} \label{Sec:obs} 

\subsection{LOFAR} \label{Sec:lofar} 
In this study, we used the International LOFAR Station LV614 (longitude: 21.8558311 deg, latitude: 57.5570617 deg)  at the Irbene Observatory (Latvia) in the stand-alone mode.
The International LOFAR Telescope is a new-generation radio interferometric array. LOFAR consists of thousands of dipole antennas distributed in 24 core stations and 16 remote stations throughout the Netherlands and 14 international stations throughout Europe \citep{van2013lofar}.
There are two distinct antenna types: Low Band Antennas (LBAs), which operate at frequencies of 10--90~MHz and High Band Antennas (HBAs), which operate at 110--240~MHz. 
The system equivalent flux density (SEFD) in a single polarization is 1.9 SFU at 45~MHz.  

\subsection{Analysed data} \label{Sec:data} 
The analysed observations were conducted between June 2022 and August 2025, with a weekly cadence, a time resolution of 1~s, and a spectral resolution of 0.195~MHz. Most observations were made in the 29--60~MHz frequency range. In all observing sessions, Cassiopeia A and the Sun were observed simultaneously. In total, 194 observational sessions were carried out, with {a} cumulative duration of 1125.2~h.

\begin{figure} 
\centerline{\includegraphics[width=1\textwidth, height=0.5\textwidth,clip=]{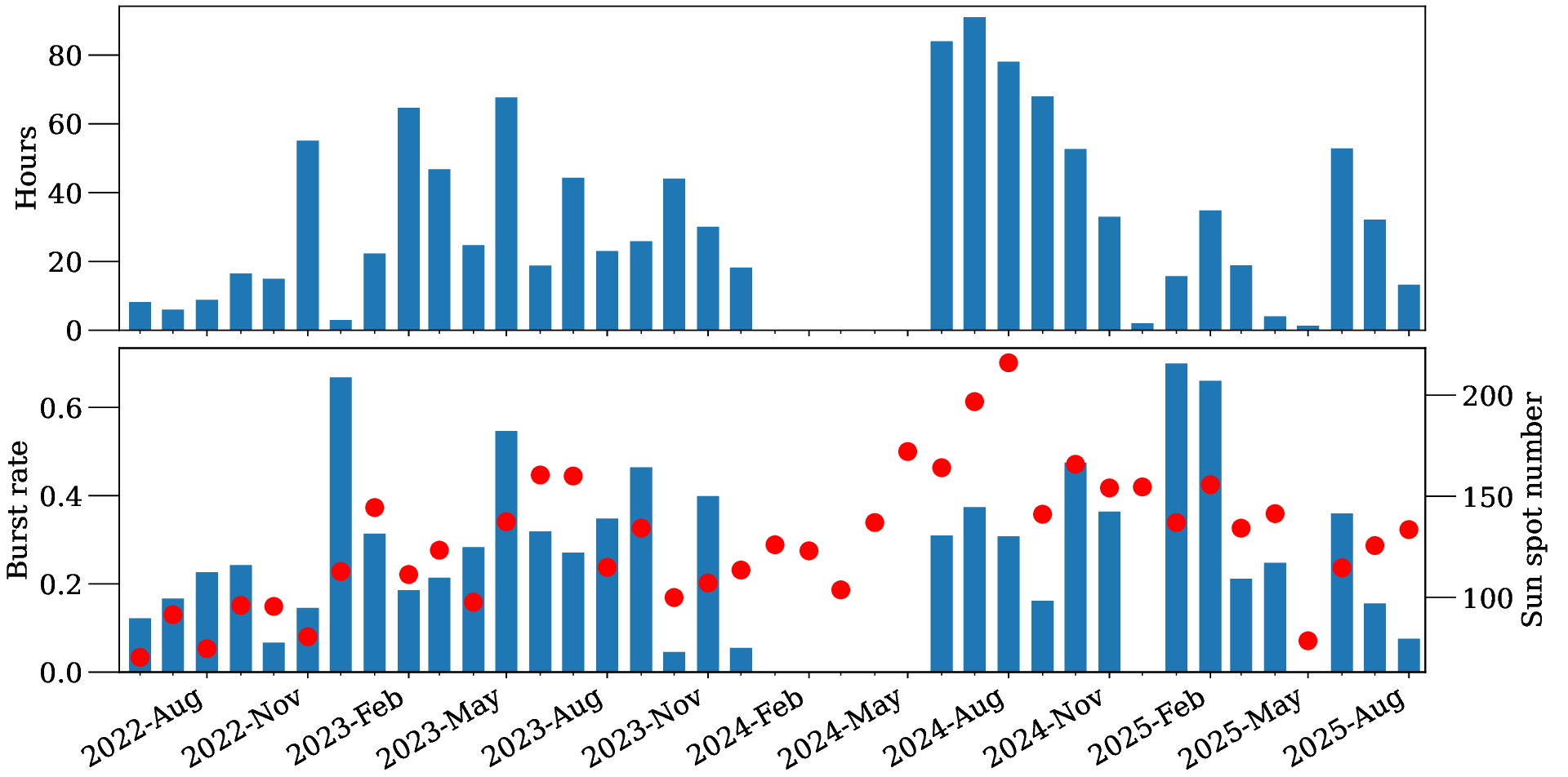}}
\caption{The number of observing hours per month (top) and the hourly rate of detected solar radio bursts (bottom). The red dots are the monthly sunspot numbers.}
\label{fig:pstasts1}
\end{figure}

During the initial phase (June 2022--September 2024), LOFAR observations were performed in two modes: as part of the International LOFAR Telescope and in single-station mode. In this period, single-station observations were conducted two days a week. In the second phase (after September 2024), International LOFAR Telescope observations were discontinued due to the LOFAR 2.0 upgrade, and only single-station observations were conducted. This observational strategy enabled the construction of a catalogue of solar radio bursts with absolute flux measurements. The main factors that affect observation cadence are the telescope availability, including technical problems, when it is not possible to run the observations, and the length of the day time at the location.  
In Figure~\ref{fig:pstasts1}, we show the number of observing hours per month and the hourly rate of detected solar radio bursts, in comparison with the monthly sunspot numbers. It is {apparent} that the hourly detection rate is correlated with solar activity. {However, our 3-year duration of data is not sufficient to study the 11-yr activity cycle. On the other hand, our 3 years of observations may be enough for detecting the manifestation of shorter-scale solar activity such as Quasi-Biennial Oscillations (QBO) in radio burst variability}
\citep[see e.g.][]{mehta2022cycle, kolotkov2015hilbert},
{which we consider as an interesting follow-up project.}

For all analysed observations, bandpass calibration was applied using Cassiopeia A as a reference source. This was performed by applying a running median filter with a window of 30 minutes to the Cassiopeia A data, thereby suppressing slow variations in the received signal power. The absolute flux densities were then derived using the sky model described in \citet{perley2017accurate}. 

{Observation data were cross-correlated with the NOAA catalogue by identifying overlapping time periods. For type II, III, IV, and V bursts, initial start and stop times were taken from the NOAA catalogue. These start and stop times were subsequently refined using flux values detected by LOFAR.
Type II, IV, and V bursts typically last several minutes, so the NOAA catalogue's time resolution is sufficient for these events. However, since the NOAA catalogue reports burst start and stop times only to minute precision, a single listed type III burst entry may actually correspond to multiple distinct bursts in the LOFAR data. In such cases, the burst with the highest flux was selected.
Because type I bursts can persist for multiple weeks, observation data were instead visually inspected to identify time periods during which type I emission was the dominant flux component. Some fainter bursts may have been missed using this approach but could potentially be identified through automated detection methods.
We plan to publish the raw data as open source, which will allow the dataset to be re-analyzed using automatic burst detection methods to identify additional solar bursts.
}

\subsection{Data calibration} \label{Sec:calib}
\begin{figure} 
\centerline{\includegraphics[width=1\textwidth]{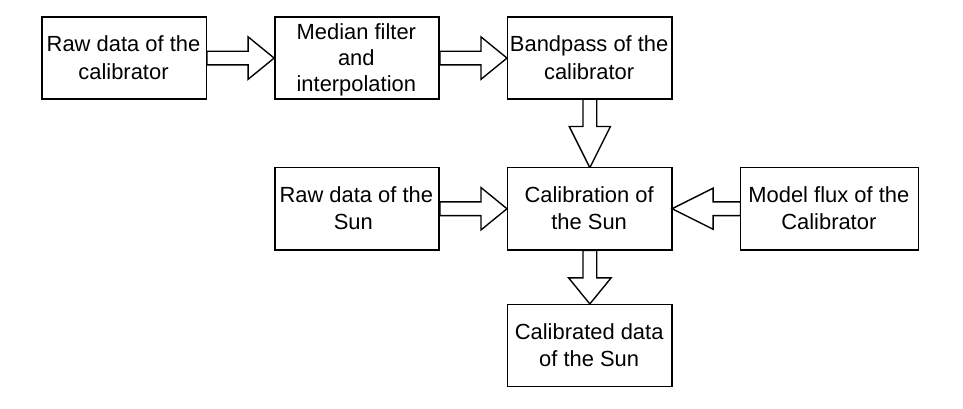}}
\caption{Calibration process diagram.}
\label{fig:calib}
\end{figure}

The calibration process of the observation of the Sun consists of the following steps. First, median filter and interpolation are applied to raw calibrator data, to estimate the bandpass of observation. Then, the estimated bandpass is multiplied by the model flux. Finally, the raw data of the Sun is divided by the output of the previous step. The overall calibration process diagram is displayed in Figure \ref{fig:calib}.

Table~\ref{tab:obs_full_list} in Appendix presents all 335 solar radio bursts detected with the LV614 LOFAR station from June 2022 to August 2025 \citep[see also][]{steinbergs_2026_20157853}.
Radio bursts of all five main types, I--V, were detected, see Section~\ref{Sec:examp}.

\subsection{Examples of solar radio bursts} \label{Sec:examp} 
\begin{figure}
        \centerline{\includegraphics[width=1\textwidth, height=0.5\textwidth]{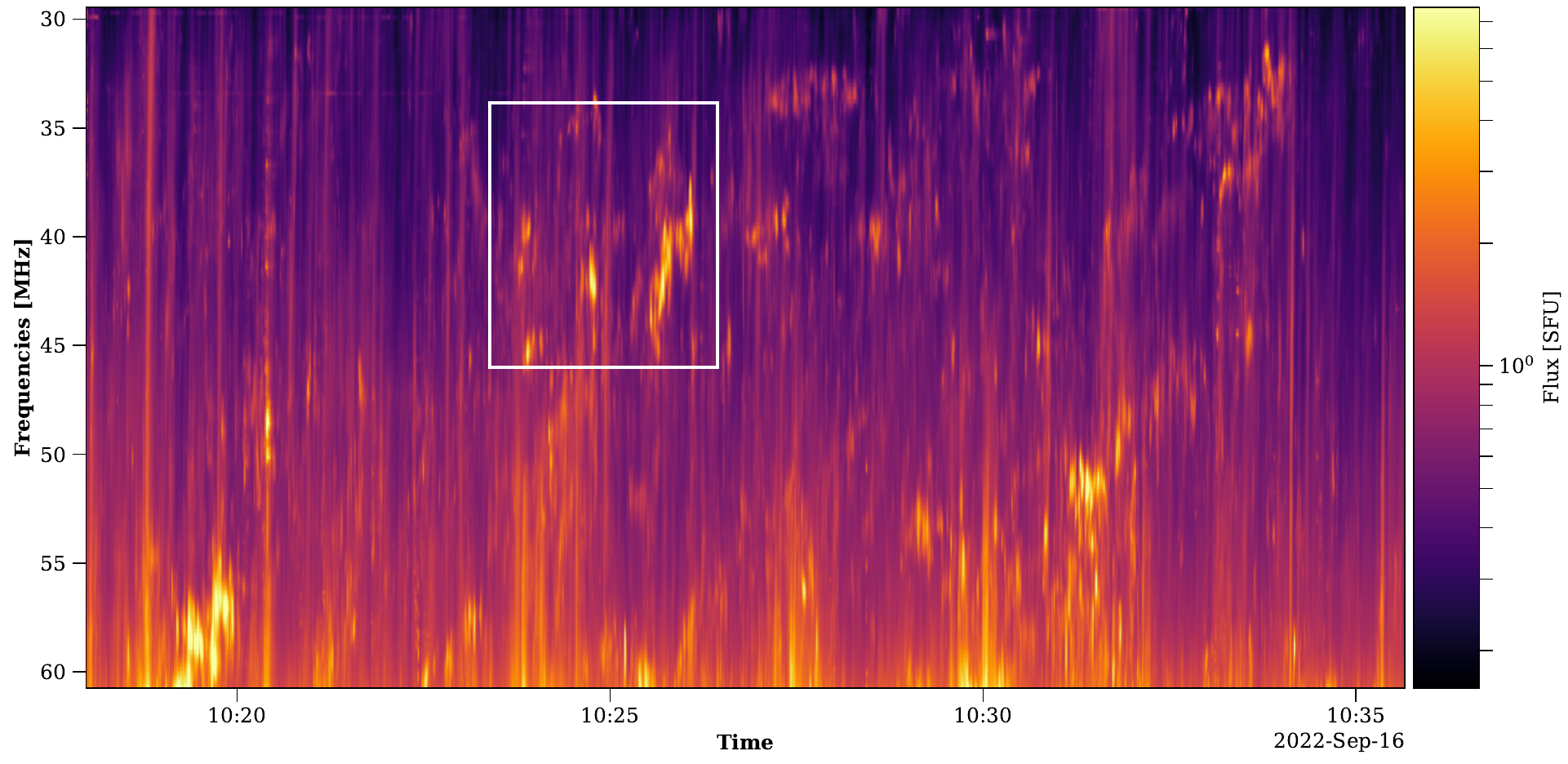}}
        \caption{Example of a Type~I solar radio burst, the white box indicates an example of patches of the burst recorded with the LOFAR station LV614. The colour scale represents the flux density (in SFU) on a logarithmic scale. The observation was performed on 2022 September 16.}
        \label{fig:b1}
\end{figure}

\noindent\textbf{\emph{Type I}.} Figure~\ref{fig:b1} shows an example of a chain of Type~I solar radio bursts recorded by the LOFAR station LV614. Type~I bursts, also known as noise storms, consist of numerous short-lived narrowband emissions superimposed on a continuum background. They typically exhibit relatively slow temporal variability compared to other types of bursts \citep{monstein2011catalog} and represent a common form of solar radio emission at metric wavelengths \citep{marassi2022trieste}. In addition, the dynamic spectrum shown in the figure contains Type~III bursts appearing as broadband lanes. In total, 92 Type~I bursts were detected in our study.

\begin{figure}
         \centerline{\includegraphics[width=1\textwidth, height=0.5\textwidth, clip=]{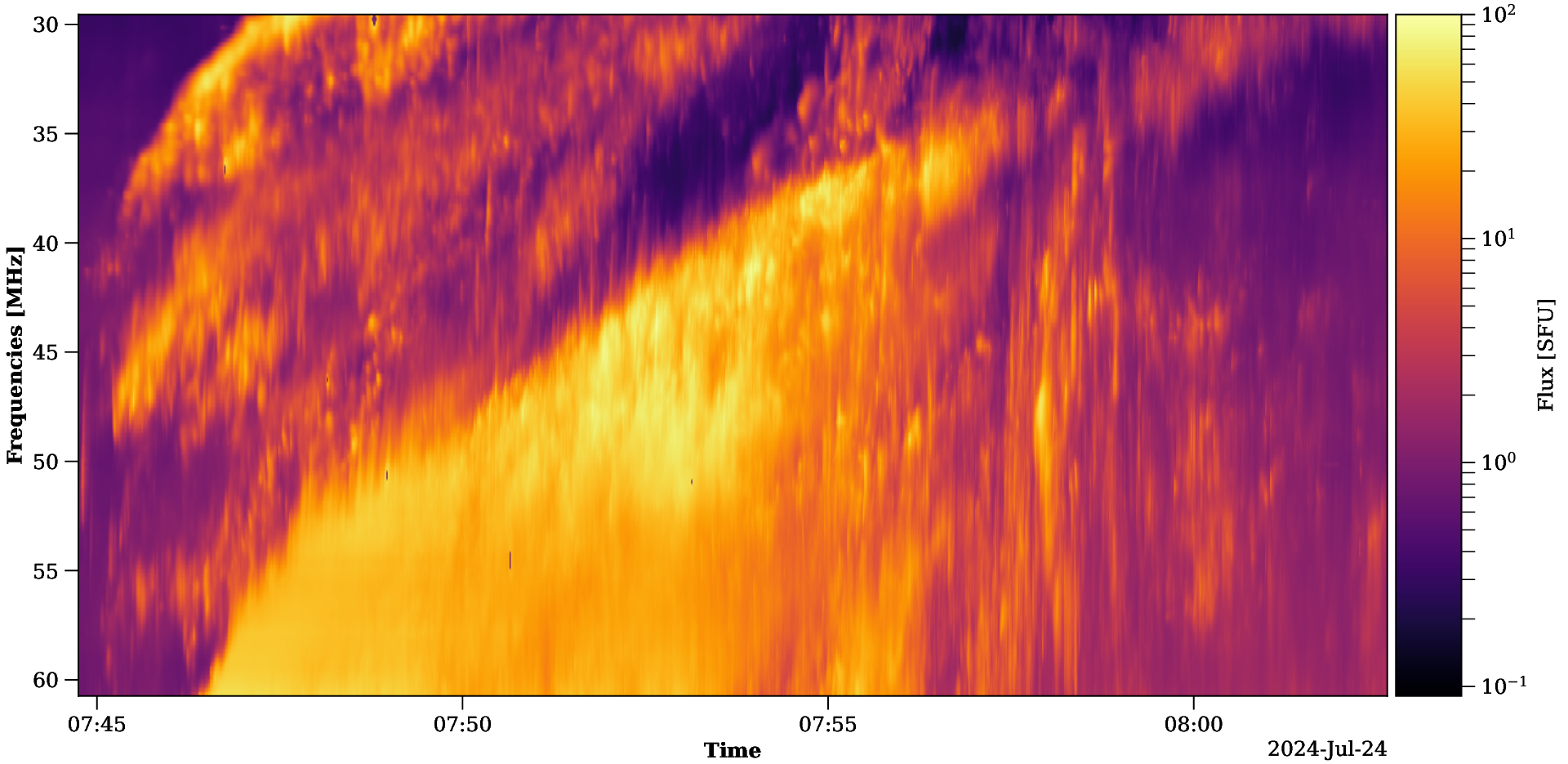}}
        \caption{Example of a Type II solar radio burst recorded with the LOFAR station LV614. The observation was performed on 2024 July 24.}
        \label{fig:b2}
\end{figure}

\noindent\textbf{\emph{Type II}.} Figure~\ref{fig:b2} shows an example of a Type~II solar radio burst recorded with LOFAR LV614. The burst exhibits both the fundamental and the second harmonic emission components. In dynamic spectra, Type~II bursts appear as slowly drifting emission bands that move from high to low frequencies, with typical  drift rates of about $-0.1~\mathrm{MHz\,s^{-1}}$ \citep{mann1996catalogue}. In total, 9 Type~II bursts were detected in our study.

\begin{figure}
         \centerline{\includegraphics[width=1\textwidth, height=0.5\textwidth, clip=]{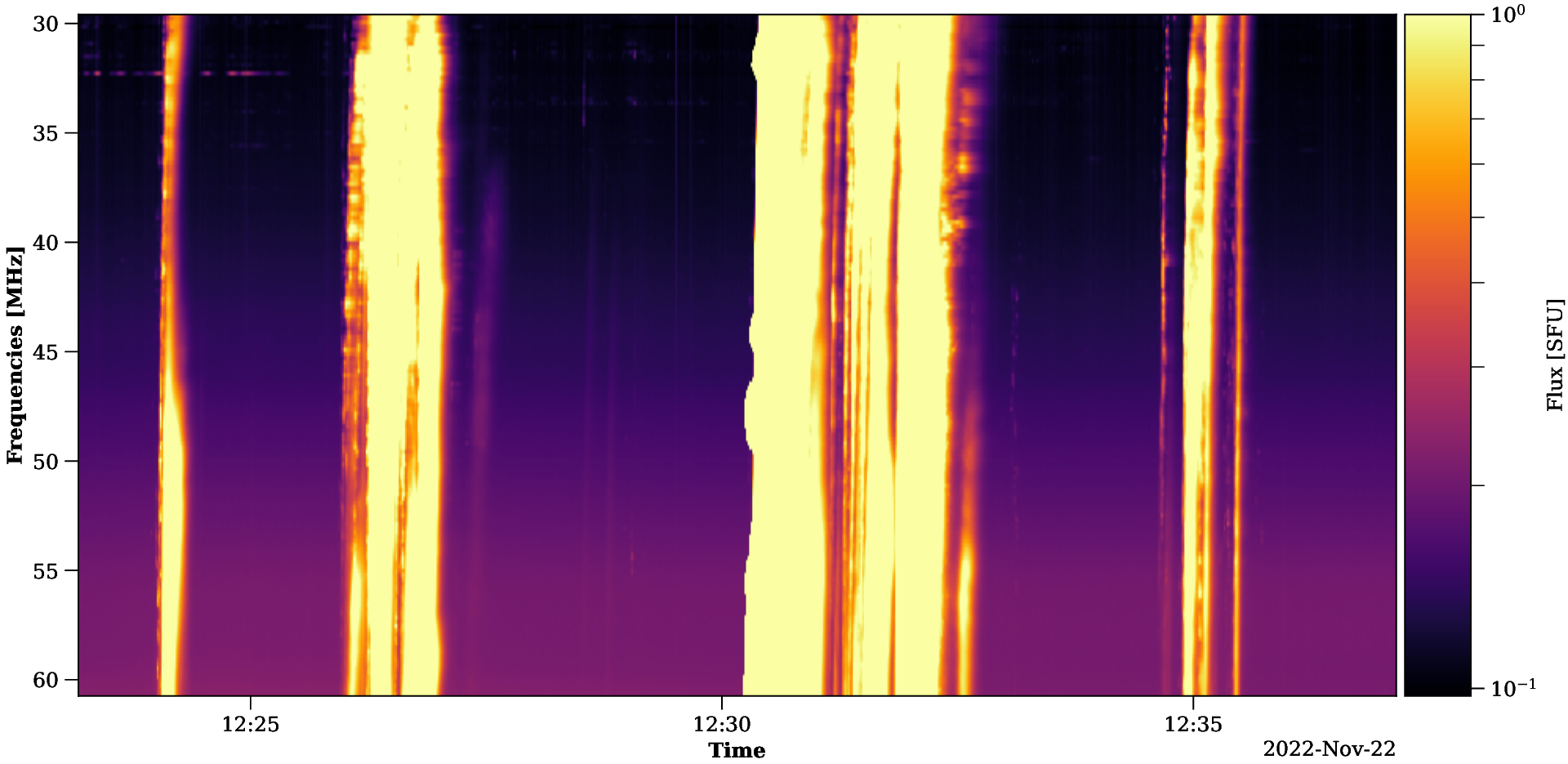}}
        \caption{Example of a Type III solar radio burst recorded with the LOFAR station LV614. The observation was performed on 2022 November 22.}
        \label{fig:b3}
\end{figure}

\noindent\textbf{\emph{Type III}.} Figure~\ref{fig:b3} shows several groups of Type~III solar radio bursts. In dynamic spectra, these bursts typically appear as nearly vertical emission lanes, reflecting their rapid frequency drift, typically about 0.1--10~MHz\,s$^{-1}$ \citep{dabrowski2021type}. In total, 298 Type~III bursts were detected in our study.

\begin{figure}
         \centerline{\includegraphics[width=1\textwidth, height=0.5\textwidth, clip=]{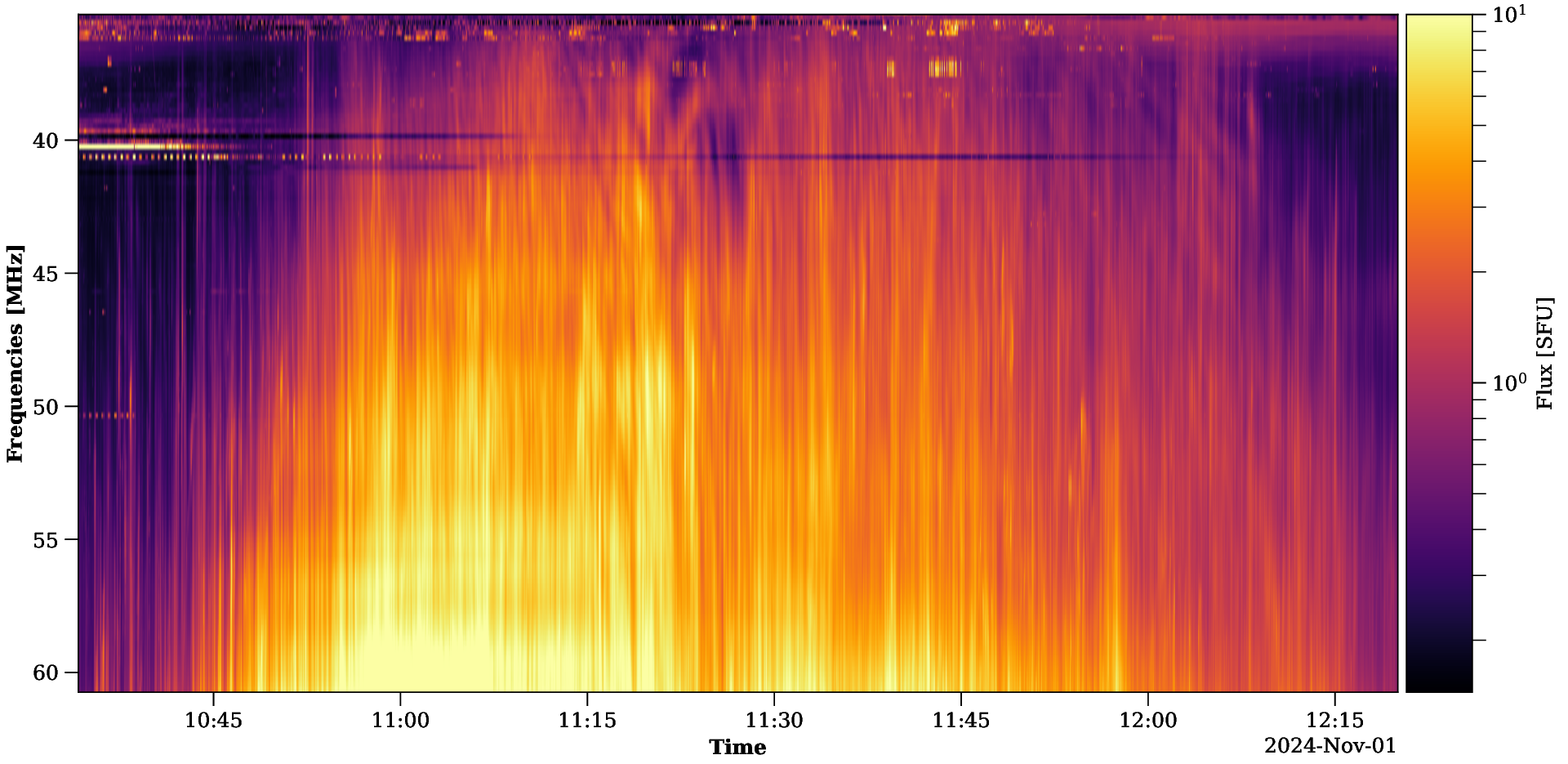}}
        \caption{Example of Type IV solar radio burst recorded with the LOFAR station LV614. The observation was performed on  2024 November 01}
        \label{fig:b4}
\end{figure}

\noindent\textbf{\emph{Type IV}.} Figure \ref{fig:b4} shows an example of Type IV  radio solar bursts. Type IV bursts have two subtypes, moving and stationary. The difference between these types is that moving Type IV bursts appear as a slowly drifting continuum toward lower frequencies, whereas stationary Type IV bursts appear as a broadband continuum fixed in a specific frequency range \citep{liu2018solar}. In this example, we see a stationary Type IV burst that lasts about 1.5~h. In total, 7 Type~IV bursts were detected in our study.

\begin{figure}
         \centerline{\includegraphics[width=1\textwidth, height=0.5\textwidth, clip=]{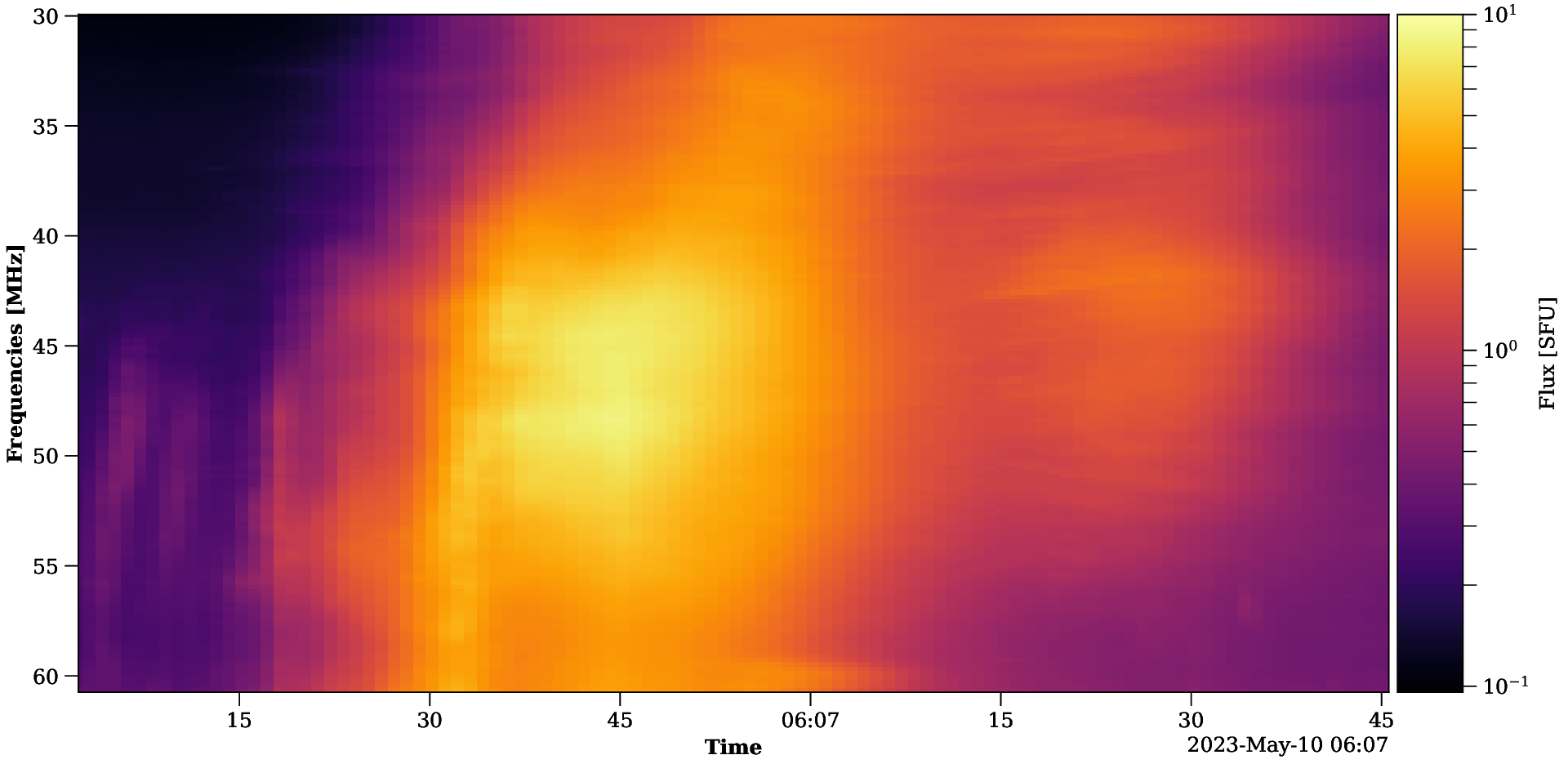}}
        \caption{Example of a solar radio type V burst recorded with the LOFAR station LV614. The Type V burst is preceded by weak Type~III bursts. The observation was performed on 2023 May 10.}
        \label{fig:b5}
\end{figure}

\noindent\textbf{\emph{Type V}.} An example of a Type~V solar radio burst is shown in Figure~\ref{fig:b5}. The Type~V emission is preceded by weak Type~III bursts. Type~V bursts are characterised by continuum emission that lasts from one to several minutes \citep{marassi2022trieste}. 
In total, 6 Type~V bursts were detected in our study.

\section{Statistical properties of detected radio bursts} \label{Sec:stat}

\subsection{Burst detection rates}

\begin{figure}
        \centerline{\includegraphics[width=1\textwidth, height=0.5\textwidth, clip=]{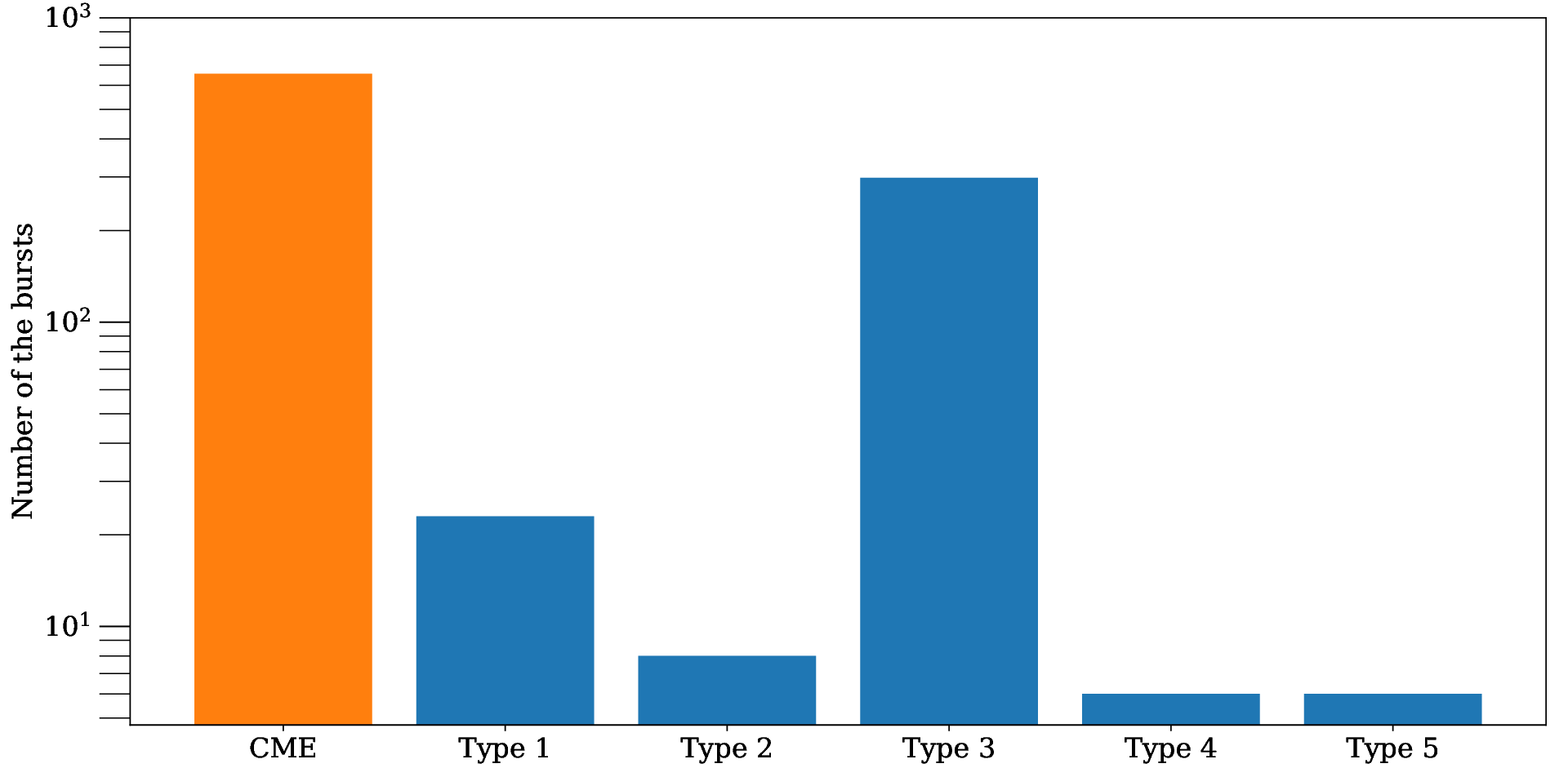}}
\caption{{Number of solar radio bursts per type.}The orange bar shows the number of detected CME (during the observations) reported by the CACTUS {catalogue}.}
        \label{fig:pstasts3}
\end{figure}

Figure~\ref{fig:pstasts3} shows the distribution of detected radio bursts by their type. For comparison, we give the number of CMEs observed in the time intervals of our observations, taken from the Computer Aided CME Tracking (CACTUS) catalogue\footnote{\url{https://www.sidc.be/cactus/}}.

The average hourly detection rates for different types of bursts are 0.02, 0.007, 0.264, 0.005, and 0.005 for Type I, II, III, IV, and V, respectively. These rates are consistent with previously reported, see Table~\ref{table:tab0}.
In particular, our detection rate for type II bursts, 0.008, is similar to the rate obtained by \citet{2021AdSpR..68.3464U}, from 0.0005 to 0.009. The latter work used data obtained during 2010--2019 at 150--450~MHz with the Compound Astronomical Low-cost Low-frequency Instrument for Spectroscopy and Transportable Observatory (CALLISTO). 
Our detection rate of type III bursts is slightly higher than the rate reported by \citet{2018A&A...618A.165Z}, from 0.02 to 0.1, who used the Nan{\c{c}}ay Decameter Array data from 2012 to 2017.  
The detection rate of type IV bursts, obtained in our study, is higher than the rate of solar decameter hectometric (1--14~MHz) type IV radio bursts which range from 0.0001 to 0.001 bursts per hour \citep{2024ApJ...971...86M}.


\begin{table}[htbp]
\begin{tabular}{||c c c c c||} 
 \hline
 Minimum hourly rate & Maximum hourly rate & Average rate & Type & Reference \\ [0.5ex] 
 \hline\hline
 0.0001 &  0.018  & 0.006 & II & [1] \\ 
 0.0006 & 0.01 & 0.004 & II & [2]\\ 
0.019 & 0.1 & - & III & [3] \\ 
0.0002 & 0.01 & 0.005 & III & [4] \\ 
0 & 0.014  & 0.004 & IV & [5] \\ 
0 &  0.001 & 0.0004& IV & [6] \\
\hline
\end{tabular}
\caption{Previously reported hourly detection rates of solar radio bursts of various types. References [1], [2], [3], [4], [5] and [6] 
correspond to \citep{2023A&A...675A.102K}, \citep{2021AdSpR..68.3464U}, \citep{2018A&A...618A.165Z}, \citep{karlsson2011local}, \citep{kumari2021occurrence}, and \citep{2024ApJ...971...86M}, respectively. }
\label{table:tab0}
\end{table}

\subsection{Peak fluxes and frequencies}
\begin{figure}
        \centerline{\includegraphics[width=1\textwidth, height=0.5\textwidth, clip=]{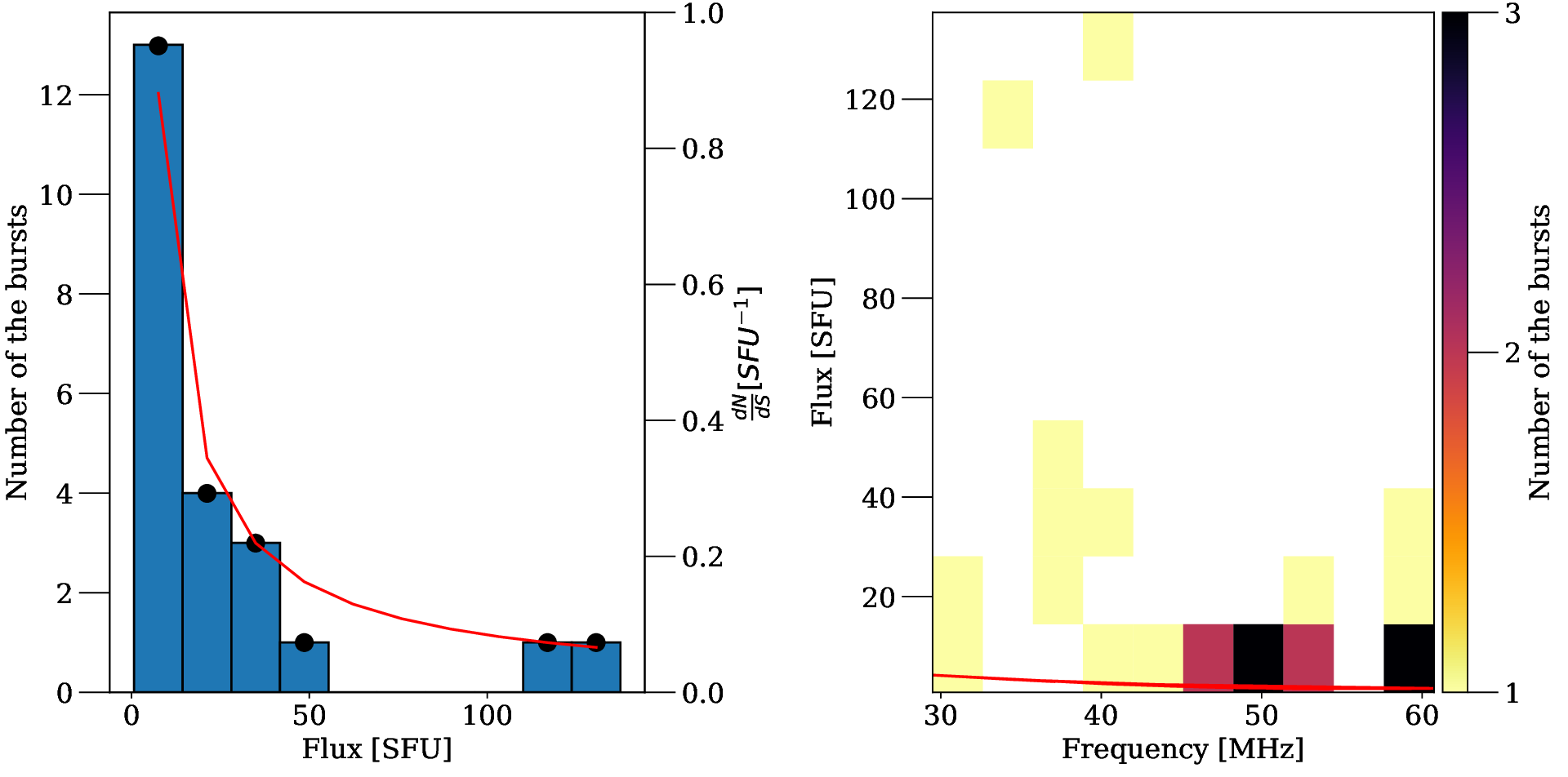}}
        \caption{Distribution of the peak fluxes of Type I solar bursts. Left: the horizontal axis shows the peak flux in SFU, and the vertical axis shows the burst count. Right: the horizontal axis shows the peak frequency in MHz, and the vertical axis shows the peak Flux in SFU. The colour scheme indicates the burst count. In the right panel, the red curve shows the system equivalent flux density (SEFD) level. In the left panel, the black circles show the number of bursts $N_i$ per each flux bin $S_i$ relative to the bin width $\Delta S$, providing  $N_i/\Delta S \approx {dN}/{dS}$. The red curve shows its best-fit by Eq.~(\ref{for:for1}).
        }
        \label{fig:pstasts4}
\end{figure}

\begin{figure}
         \centerline{\includegraphics[width=1\textwidth, height=0.5\textwidth, clip=]{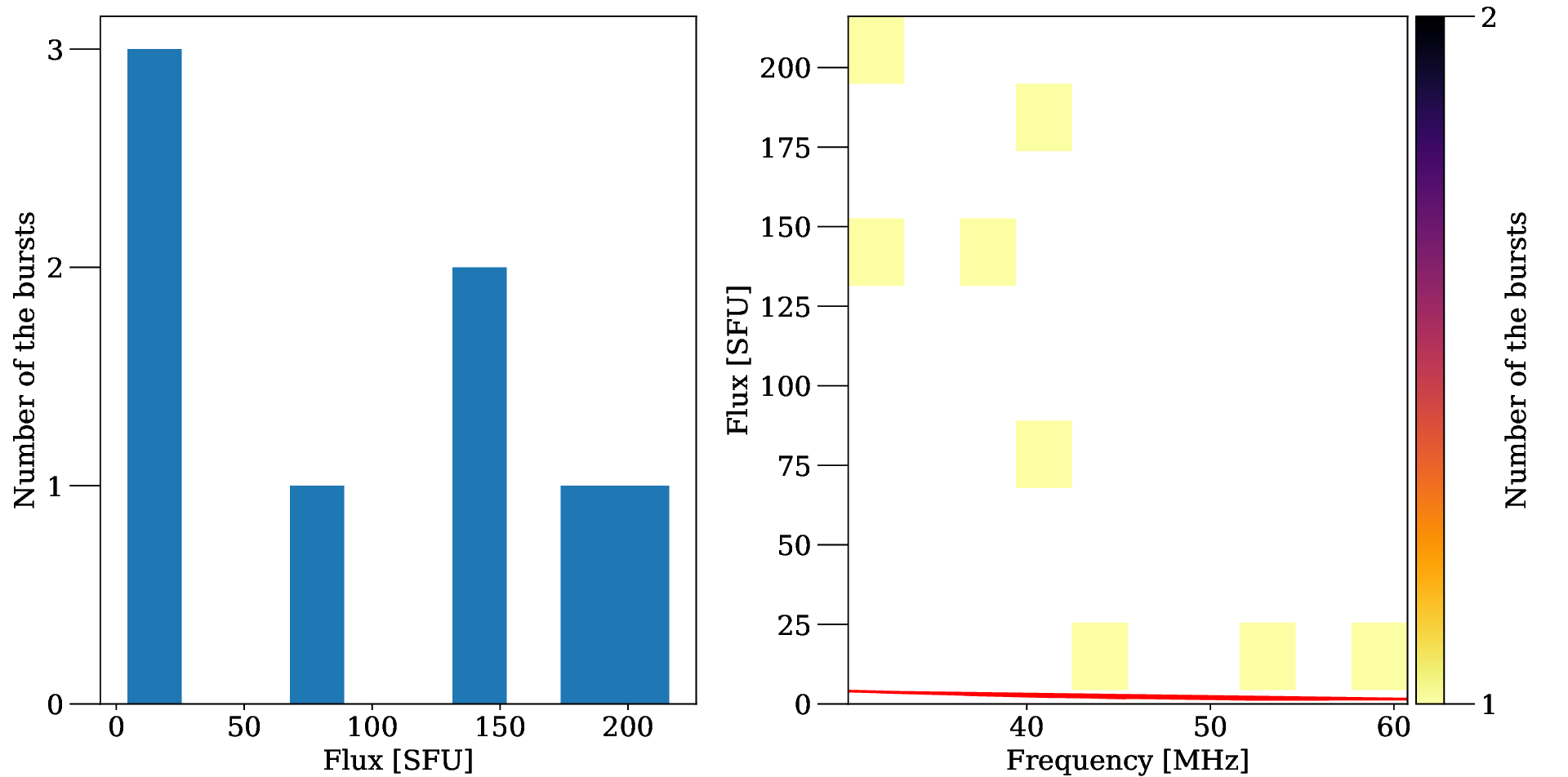}}
        \caption{The same as in Figure~\ref{fig:pstasts4} but for Type II bursts and without the power-law fit. }
        \label{fig:pstasts5}
\end{figure}

\begin{figure}
         \centerline{\includegraphics[width=1\textwidth, height=0.5\textwidth, clip=]{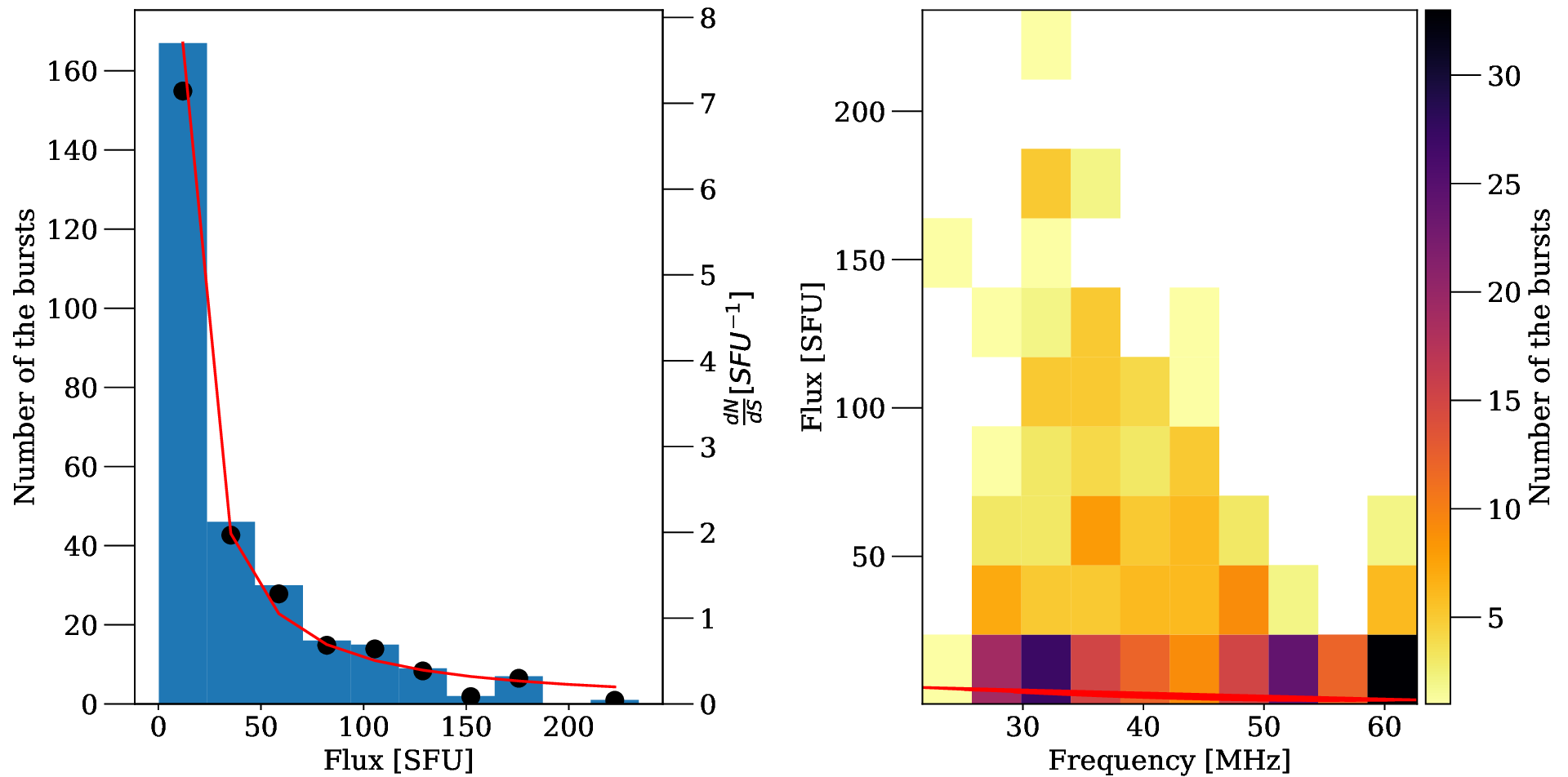}}
        \caption{The same as in Figure~\ref{fig:pstasts4} but for Type III bursts.}
        \label{fig:pstasts6}
\end{figure}

\begin{figure}
         \centerline{\includegraphics[width=1\textwidth, height=0.5\textwidth, clip=]{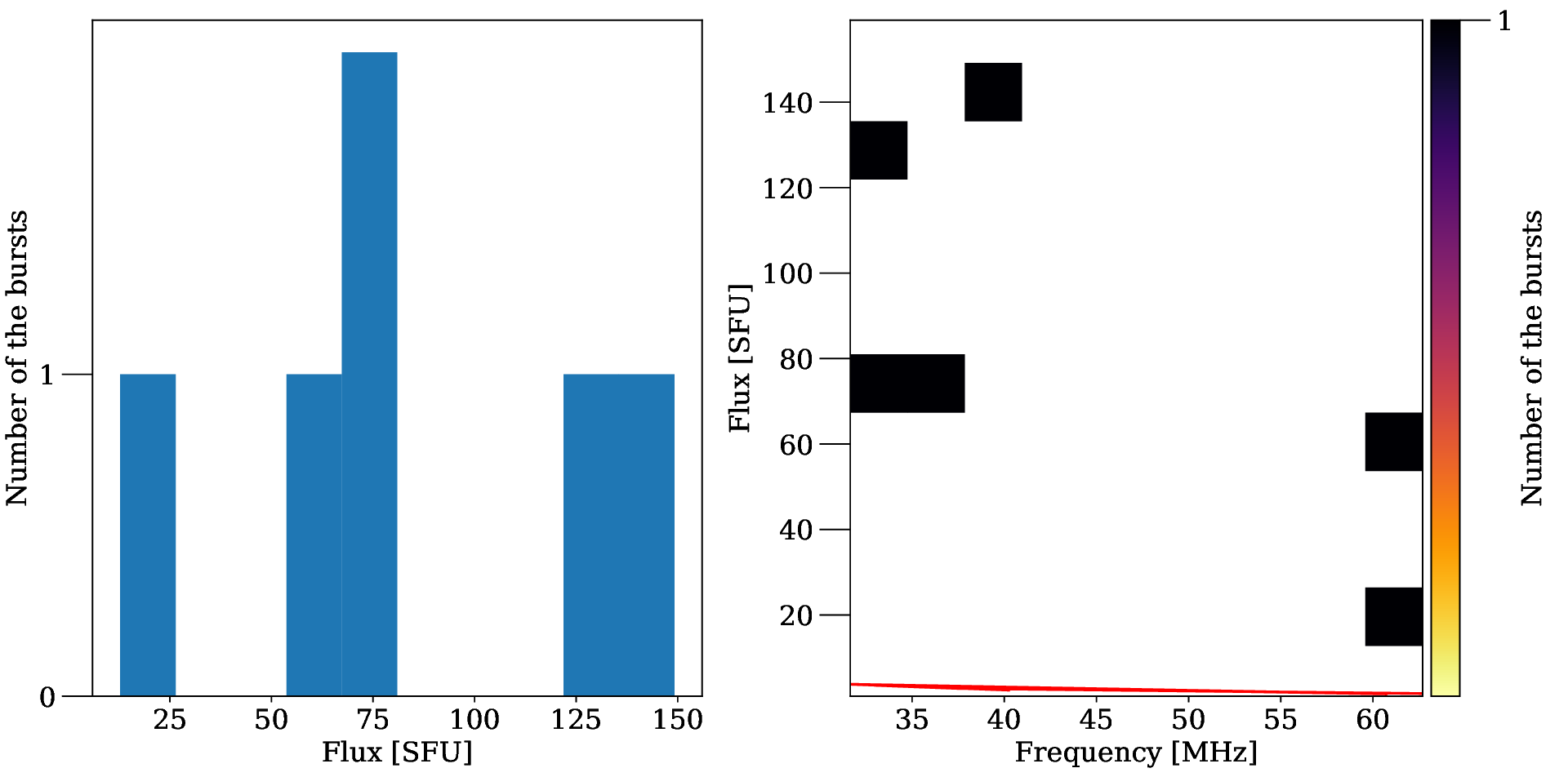}}
        \caption{The same as in Figure~\ref{fig:pstasts4} but for Type IV bursts and without the power-law fit.}
        \label{fig:pstasts7}
\end{figure}

\begin{figure}
         \centerline{\includegraphics[width=1\textwidth, height=0.5\textwidth, clip=]{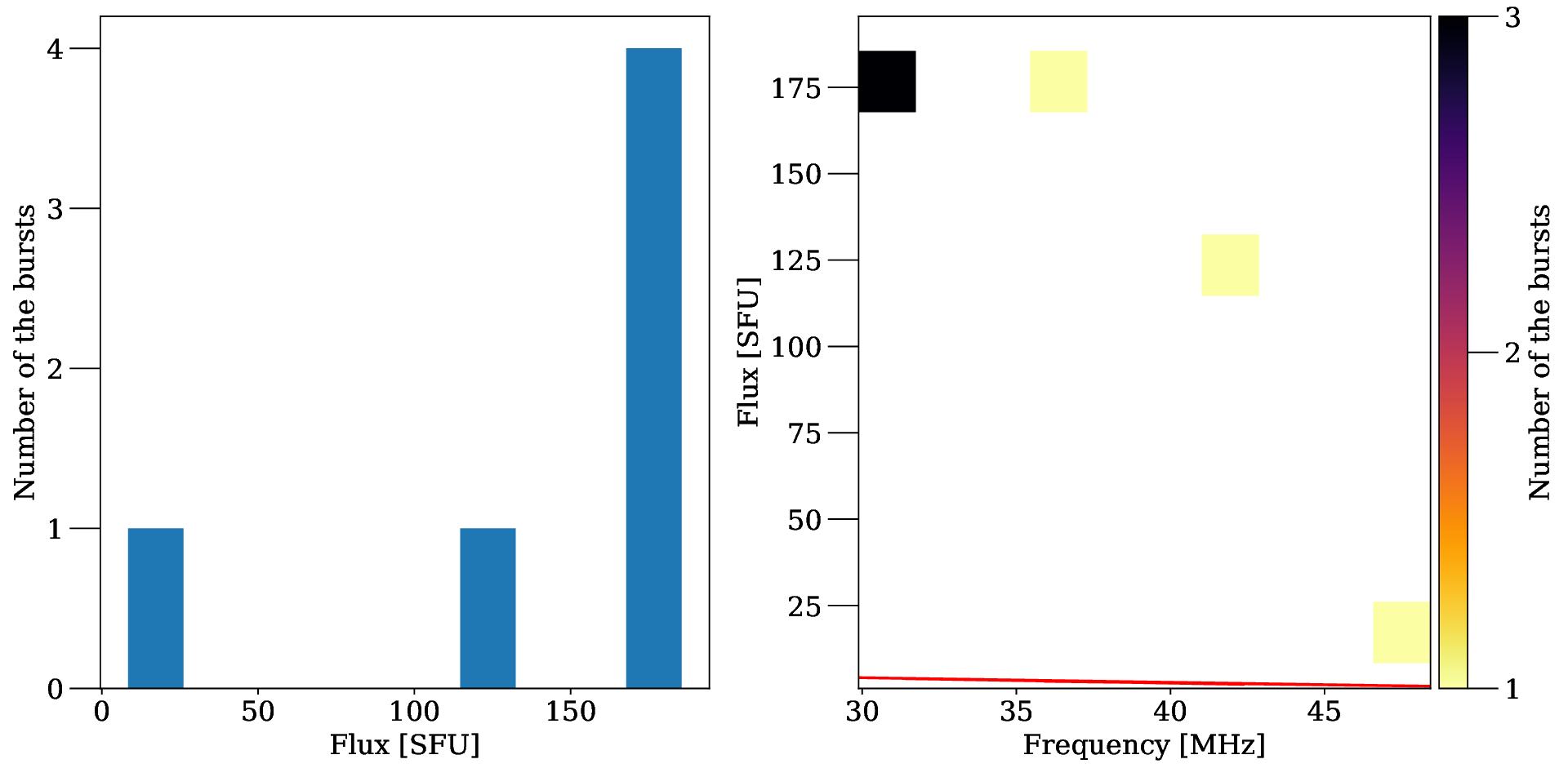}}
        \caption{The same as in Figure~\ref{fig:pstasts4} but for Type V bursts and without the power-law fit.}
        \label{fig:pstasts8}
\end{figure}

Parameters of the detected radio bursts allow us to construct a quantitative overview of the distribution of both peak fluxes and frequencies for bursts of various types.
Figures~\ref{fig:pstasts4}--\ref{fig:pstasts8} show statistics of peak fluxes and peak frequencies for Types I, II, III, IV, and V, detected in our study, respectively. In each figure, the histogram of the number of events with the peak flux (left panel)reveals the dependence of the occurrence rate of the burst population on the burst strength, which can indicate differences in emission efficiency and source region energetics.

The 2D flux–frequency distributions (right panels in Figs.~\ref{fig:pstasts4}--\ref{fig:pstasts8}) highlight how the burst strength varies with the peak emission frequency. Because the radio emission frequency corresponds directly to the local electron concentration in the emission source, systematic trends, such as stronger bursts occurring preferentially at specific frequencies, can reveal differences in source height, particle acceleration mechanisms, and emission regimes. Comparing these patterns across burst types enables the identification of distinct populations and helps constrain which physical processes dominate in different coronal environments.
In particular, Type III events, which are the most statistically numerous (see Fig.~\ref{fig:pstasts6}), show a clear dependence of flux on frequency. Low-flux bursts (below 50\,SFU) are distributed over a relatively wide frequency range of 30--60\,MHz, whereas more energetic bursts (above 100\,SFU) are predominantly confined to a narrower band, around 30--40\,MHz. 

{The right-hand panels in Figs.~\ref{fig:pstasts4}--\ref{fig:pstasts8} also show the corresponding SEFD levels, illustrating the frequency dependence of the instrumental sensitivity. LOFAR single international stations have these sensitivity values at 41 kJy, 19 kJy, and 15 kJy for the 30 MHz, 45 MHz, and 60 MHz} \citep{van2013lofar}, {respectively.
The frequency-dependent sensitivity of LOFAR is also consistent with the general behaviour expected for low-frequency radio interferometric observations.
For comparison, Long Wavelength Array Station (LWA) has 13.7 kJy, 10.6 kJy, 10.3 kJy, and 11.5 kJy for the 76, 60, 44, and 28 MHz passbands, respectively }\citep{ellingson2013observations}.

\subsection{Peak flux densities}
\label{s:peak_flux_dens}

\citet{1997ApJ...474L..65M, suresh2017wavelet} established that the decrease in the number of bursts with the peak flux density $S$, follows a power law,
\begin{equation}\label{for:for1}
\frac{dN}{dS} = AS^{\alpha}, 
\end{equation}
where $S$ is the peak flux of the burst, $N$ is the number of bursts, and the power-law index $\alpha$ is constant.
We apply Eq.~(\ref{for:for1}) to the histograms shown in Figs.~\ref{fig:pstasts4} and \ref{fig:pstasts6} for type I and type III events, respectively, as they constitute the most statistically well-sampled populations. In these histograms, we extract the number of bursts $N_i$ per each flux bin vs. the corresponding bin centres $S_i$, so that ${dN}/{dS} \approx N_i/\Delta S$, where $\Delta S$ is the bin width (13.67\,SFU for type I and 23.38\,SFU for type III events).
Parameters $A$ and $\alpha$ are estimated using a least absolute deviations regression method known to be less sensitive to outliers than the standard least-squares techniques.
The best fitting values obtained for type I and type III solar bursts with the corresponding 1-$\sigma$ uncertainties are shown in Table ~\ref{table:tab2} and indicated in Figures~\ref{fig:pstasts4} and \ref{fig:pstasts6}.





%
%
%

\begin{table}
\begin{tabular}{|| c c c c c ||} 
\hline
Solar burst type & $\alpha$ & 1-$\sigma$ error($\alpha$) & $A$ & 1-$\sigma$ error($A$)  \\ [0.5ex] 
\hline\hline


I & $-0.9$ & 0.2 & 5.5 & 2.7 \\
III & $-1.25$ & 0.07 & 171 & 52 \\

\hline\end{tabular}
\caption{Parameters of the power-law fits of the distribution of type I and III bursts with the peak flux.}
\label{table:tab2}
\end{table}


{The frequency-dependent sensitivity of LOFAR and the resulting detection threshold can affect the observed peak-flux distribution, particularly at low flux densities, and may therefore influence the derived power-law slopes.
The frequency dependence of the instrumental sensitivity is illustrated in Figs.~\ref{fig:pstasts4}--\ref{fig:pstasts8}. These variations were considered during the calibration of the observations.
Thus, the instrumental sensitivity improves toward higher frequencies within the frequency range considered. Consequently, a burst with the same intrinsic flux density can have a different probability of detection depending on its observing frequency.

It is important to distinguish this instrumental sensitivity correction from the completeness of the burst catalogue. The first step of our calibration procedure is bandpass calibration, which removes the frequency-dependent instrumental response and normalizes the measured amplitudes around 1.
Absolute flux calibration is then performed using the model flux (estimated from VLA observations)}\citep{perley2017accurate}{, in our case Cassiopeia A. After these calibration steps, the measured burst flux densities are placed on an absolute flux-density scale, allowing fluxes at different frequencies to be compared. However, calibration itself does not remove the observational selection effect associated with the minimum signal-to-noise ratio required for a burst to be detected.

}

\subsection{Heights of the radio source}

The local plasma frequency ($\nu_\mathrm{pe}$) is linked with the electron concentrations ($n_\mathrm{e}$) as
\begin{equation}\label{for:for2}
  \nu_\mathrm{pe} \approx C\times s \times (n_\mathrm{e})^{1/2},
\end{equation}
where the constant $C$ is 8,980~Hz~cm$^{{3}/{2}}$, and $s$ is the harmonic number 
\citep[e.g.,][]{morosan2014lofar, mann2018radio}. 

 \begin{figure}[t!]
    \centering
        \includegraphics[width=1\textwidth]{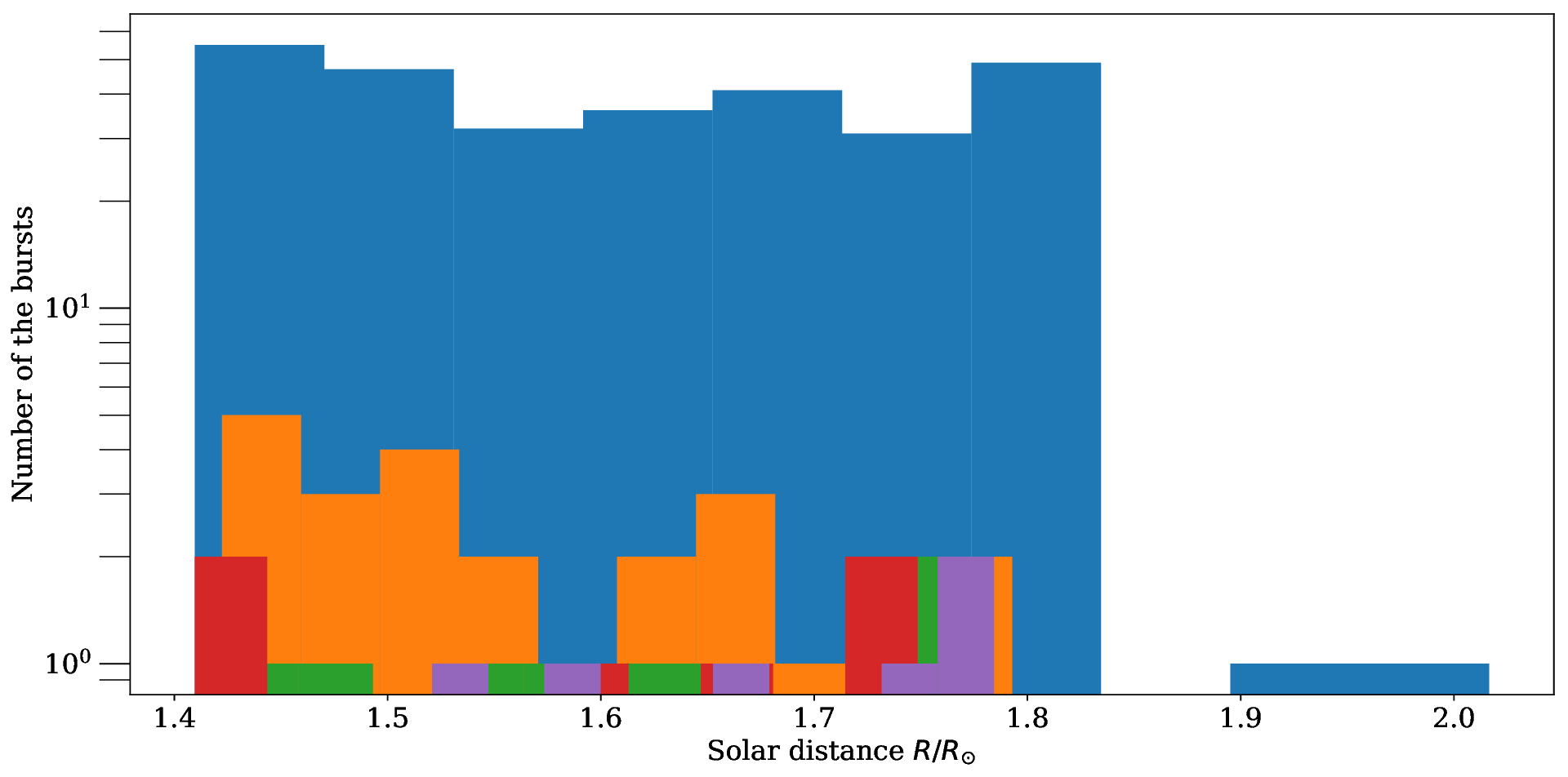}
        \includegraphics[width=1\textwidth]{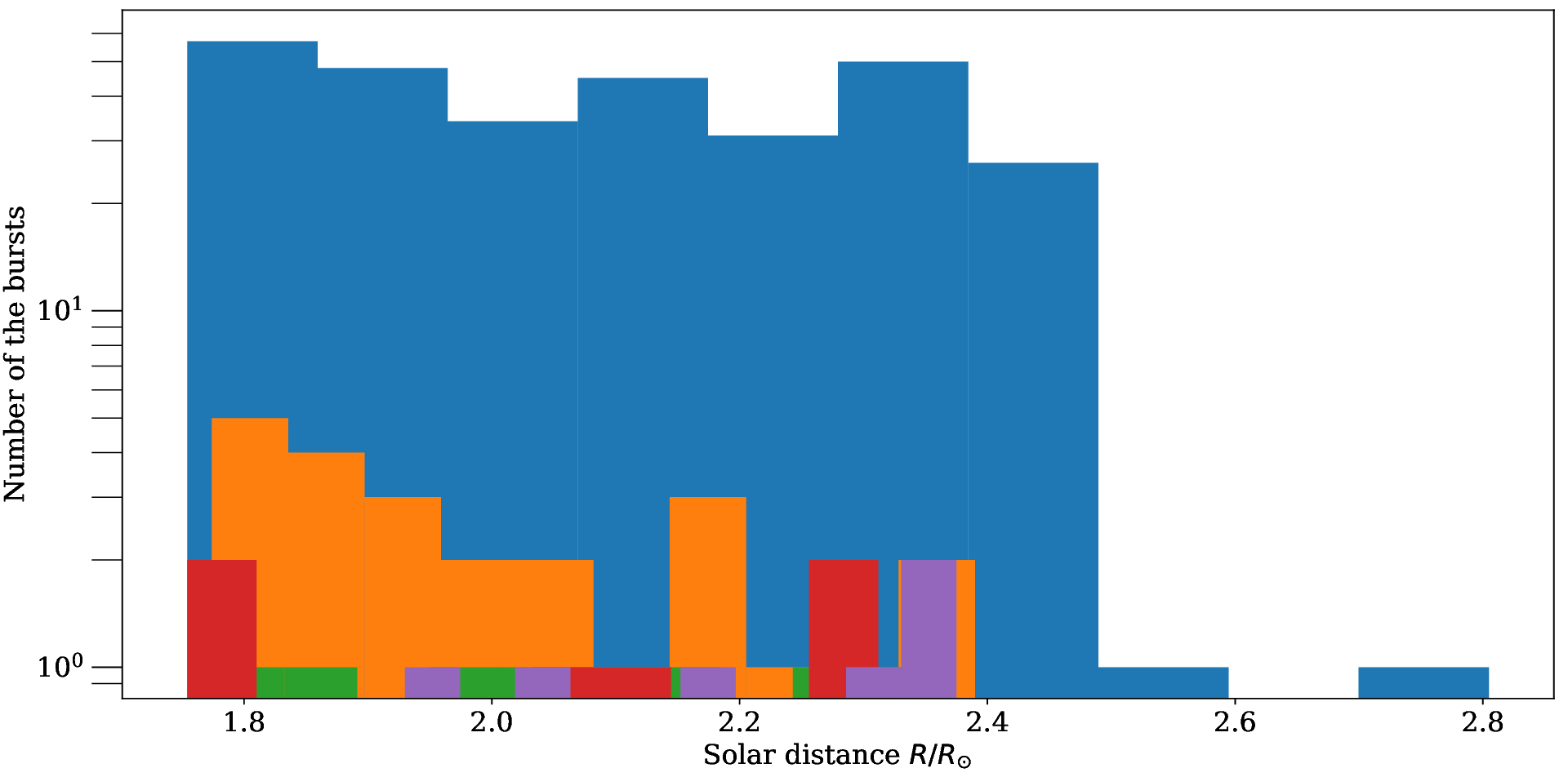}
    \caption{Distances of the radio sources, different types of solar radio bursts, estimated by the peak frequencies. The distance $R$ is shown in solar radii from the solar centre. {The distance $R$ was estimated using the Newkirk density model. Top: harmonic index ($s$) equals 1, see Eq.~(\ref{for:for2}). Bottom: harmonic index ($s$) equals 2, see Eq.~(\ref{for:for2}). The colour scheme:  type III (blue), type I (orange), type III (green),  type IV (red), type V (violet).}}
    \label{fig:r}
\end{figure}

As the frequency $f$ of the fundamental radio emission produced by the plasma mechanism equals the local plasma frequency, $f = \nu_\mathrm{pe}$ \citep{melnik2015decameter}, we can estimate the electron density in the emitting plasma volume. {In our work, we use the Newkirk coronal density model.}
It allows us to estimate the height of the radio source using the dependence of the plasma density on height {\citep{1961ApJ...133..983N}},  
\begin{equation}\label{for:for3}
  n_\mathrm{e} \approx n_{0} \times 10^{4.32 \times (R_{\odot}/{R})},
\end{equation}
where $R$ is the distance from the solar centre, $R_{\odot}$ is the solar radius, and $n_{0}$ is $4.2 \times 10^4$~cm$^{-3}$ \citep[see also][]{2021SoPh..296...57K}.
{We choose this model because of its simplicity, and we do not to intend to compare different models and/or evaluate their correctness. On the other hand, different coronal density models are not expected to alter the relative height hierarchy between solar burst types (our main scope), as shown in} \citet{morosan2014lofar}.
Figure~\ref{fig:r} shows the distances of the radio sources, estimated by the peak frequency. 
Type III bursts exhibit the broadest height distribution and dominate statistically over the entire range, while the other burst types tend to cluster within narrower intervals of heliocentric distances (may be caused by fewer detections). The presence of several Type III events at larger heights, approaching $2R_{\odot}$, is consistent with the propagation of electron beams along extended open magnetic field structures into the upper corona.

\subsection{Frequency drifts}
\begin{figure}[t!]
    \centering
        \centering
        \includegraphics[width=1\textwidth]{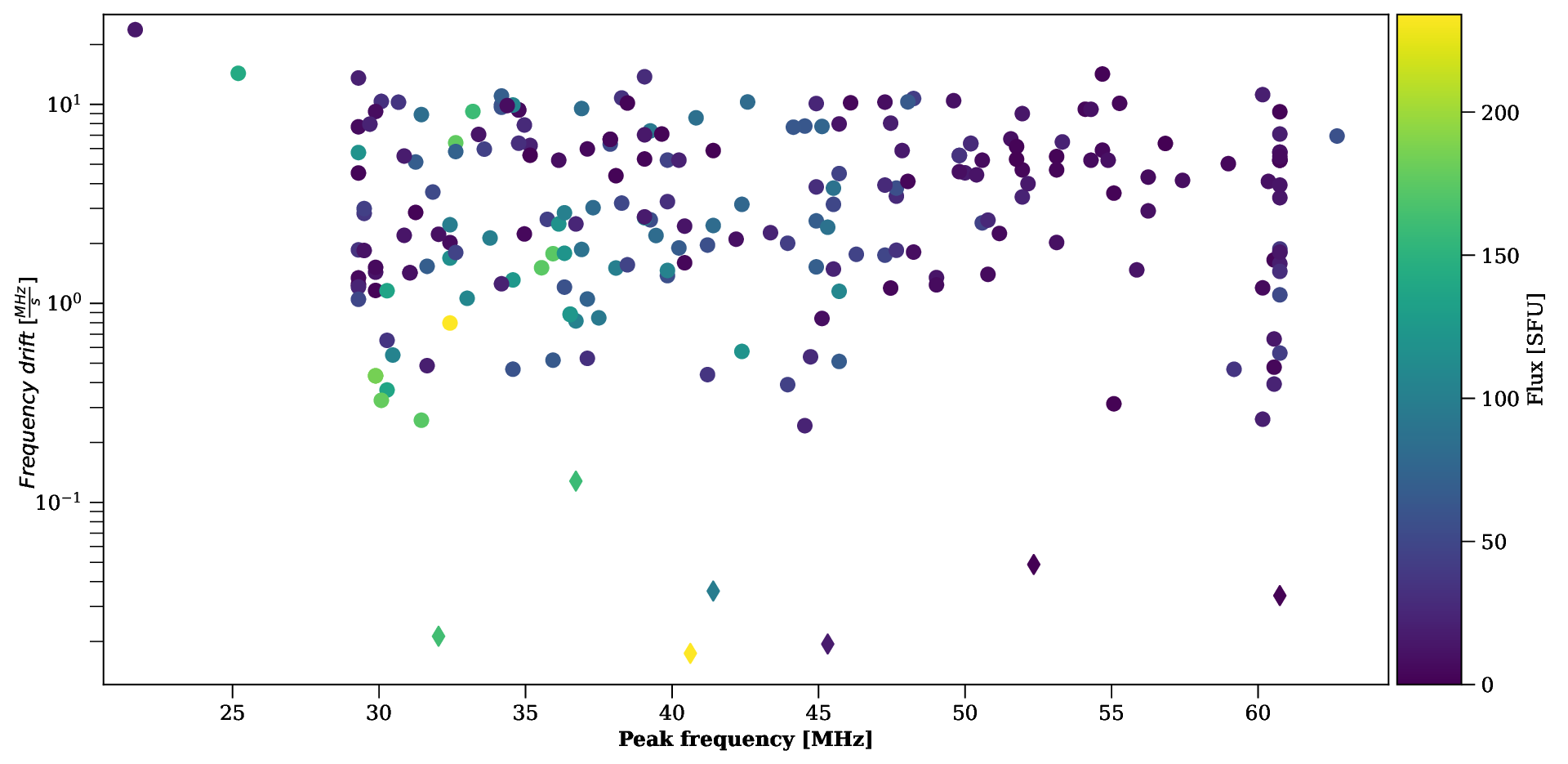}
        \label{fig:dfdt1}

        \centering
        \includegraphics[width=1\textwidth]{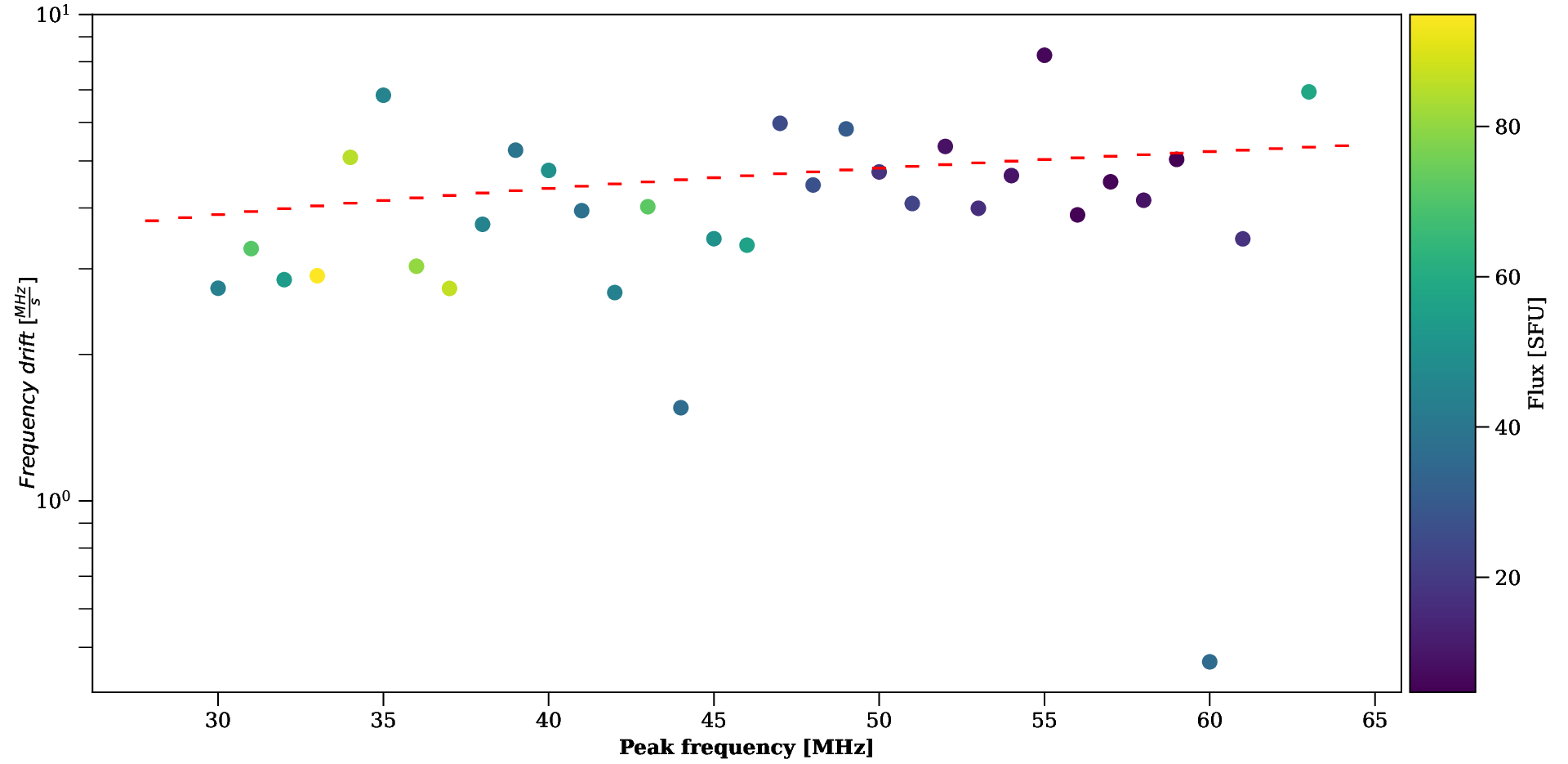}
         \label{fig:dfdt2}
    \caption{{Instantaneous} frequency drifts estimated for type III bursts (dots) and type II bursts (diamonds). The horizontal axis shows the peak frequency in MHz, and the vertical axis shows the {instantaneous} frequency drift in MHz\,s$^{-1}$. The colour scheme indicates the peak flux in SFU.  {Top: Raw data. Bottom: binned data (over 1~MHz bins) and its best-fit $df/dt = Af^a$ (red dashed line), with $A=0.78 \pm 0.91$  and $a=0.43 \pm 0.31$; these parameters were found using the least-squares method.}}
    \label{fig:dfdt}
\end{figure}

In a radio burst produced by the plasma mechanism, the frequency drift, which can be quantified by the time derivative, ${df}/{dt}$, indicates the direction and speed of propagation of the electron beam. We estimated the {instantaneous (at the moment of peak flux)} frequency drift using the following procedure. First, the instant of time at which the burst reaches its maximum flux is determined. Then, assuming a linear dependence of the instantaneous frequency on time in the vicinity of the peak, we estimate its gradient using linear regression. Figure~\ref{fig:dfdt} and Table~\ref{tab:obs_full_list} show the estimated {instantaneous} frequency drifts for Type II and III bursts.


For type II bursts, frequency drift rates obtained in our study are consistent with previous estimations \citep[see, e.g.,][]{mann1996catalogue, 2018A&A...618A.165Z, ndacyayisenga2023overview}, and have a mean value around $0.043 ~\mathrm{MHz\,s^{-1}}$.
Our estimates of the frequency drift rates for type III solar bursts are consistent with the results of \citet{reid2018solar}, who found that the drift rates at frequencies below 50\, MHz are typically smaller than 10~MHz\,s$^{-1}$. In particular, the mean type III {instantaneous} drift rate obtained in our study is $ 4.383 ~\mathrm{MHz\,s^{-1}}$.

{We compute a frequency drift fit, $df/dt = Af^{a}$, as seen in Figure \ref{fig:dfdt} (bottom panel). Our best fit parameters are $A=0.78 \pm 0.91$  and $a=0.43 \pm 0.31$. In previous studies, the corresponding best fit results were reported as $df/dt={-}0.01f^{1.84}$} \citep[][50 kHz to 3 MHz]{alvarez1973solar}, {$df/dt = 0.1f^{1.4}$ and $df/dt = 0.09f^{1.35}$} \citep{aschwanden1995solar} {and} \citep{melednez1999statistics} {respectively (100 MHz and 3000 MHz). To compute the best fit parameters, we binned our frequency drift vs. peak frequency scatter plot over 1~MHz bins. For each bin, mean frequency drift values and flux value were computed. Then the least squares method was used to compute the best fit parameters. To improve the fit, mean flux was used as fitting weights.}

\subsection{Electron beam speeds}

The electron density decreases with altitude in the corona and, therefore, the frequency also decreases with altitude. It is possible to estimate the radial velocity of the Type III bursts using the observed frequency drift rate, $df/dt$, and the density model $n_e(R)$ given by Eq.~(\ref{for:for3}), as \citep[see e.g.][]{morosan2014lofar},
\begin{equation}\label{for:for4}
   v = \frac{2 \sqrt{n_{e}}}{C} \left({\frac{dn_{e}} {dR}}\right)^{-1} \frac{df}{dt},
\end{equation}
where the constant $C$ is given in Eq.~(\ref{for:for2}).

  \begin{figure}[t!]
    \centering
        \includegraphics[width=0.8\textwidth]{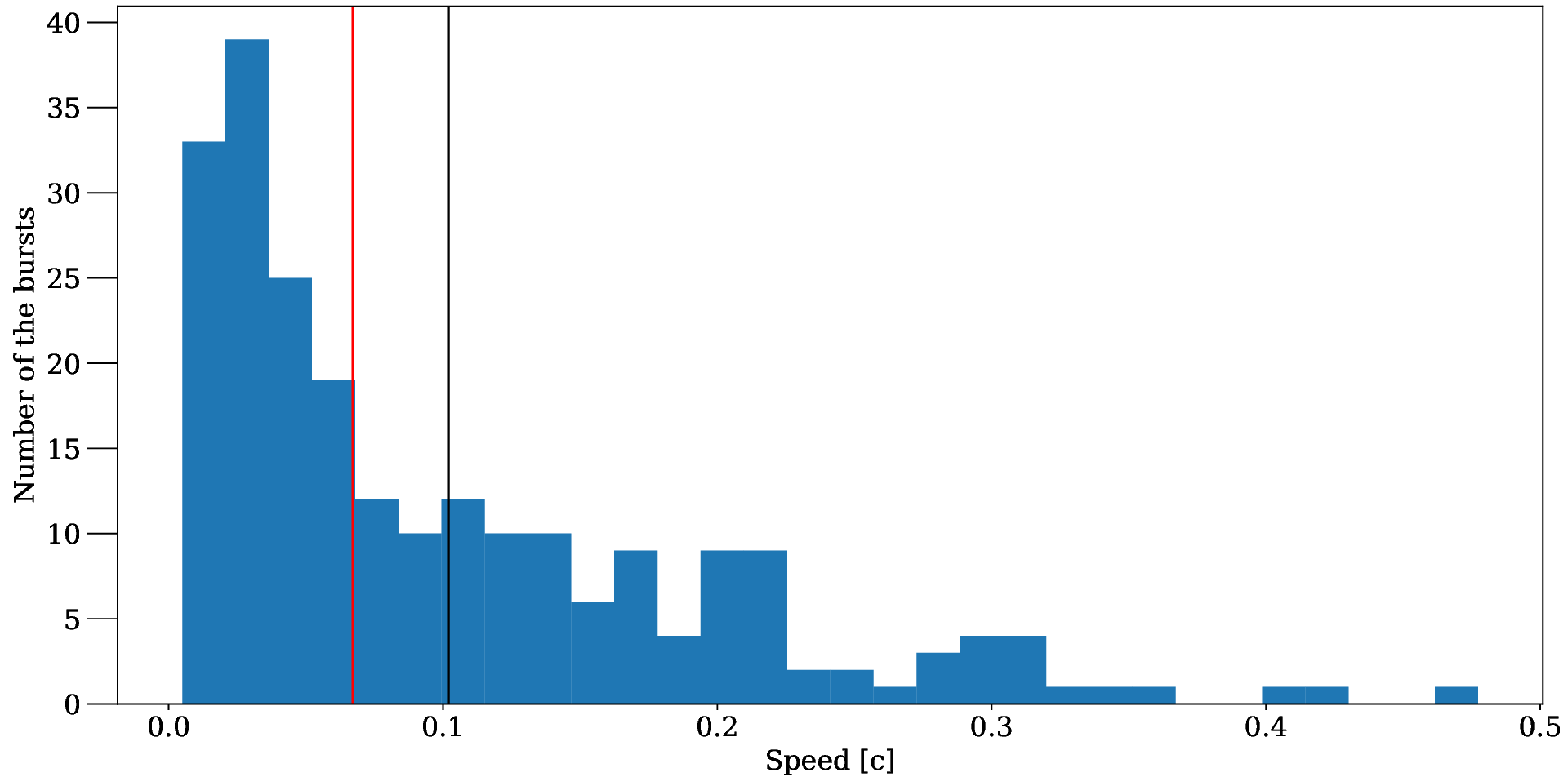}
        \includegraphics[width=0.8\textwidth]{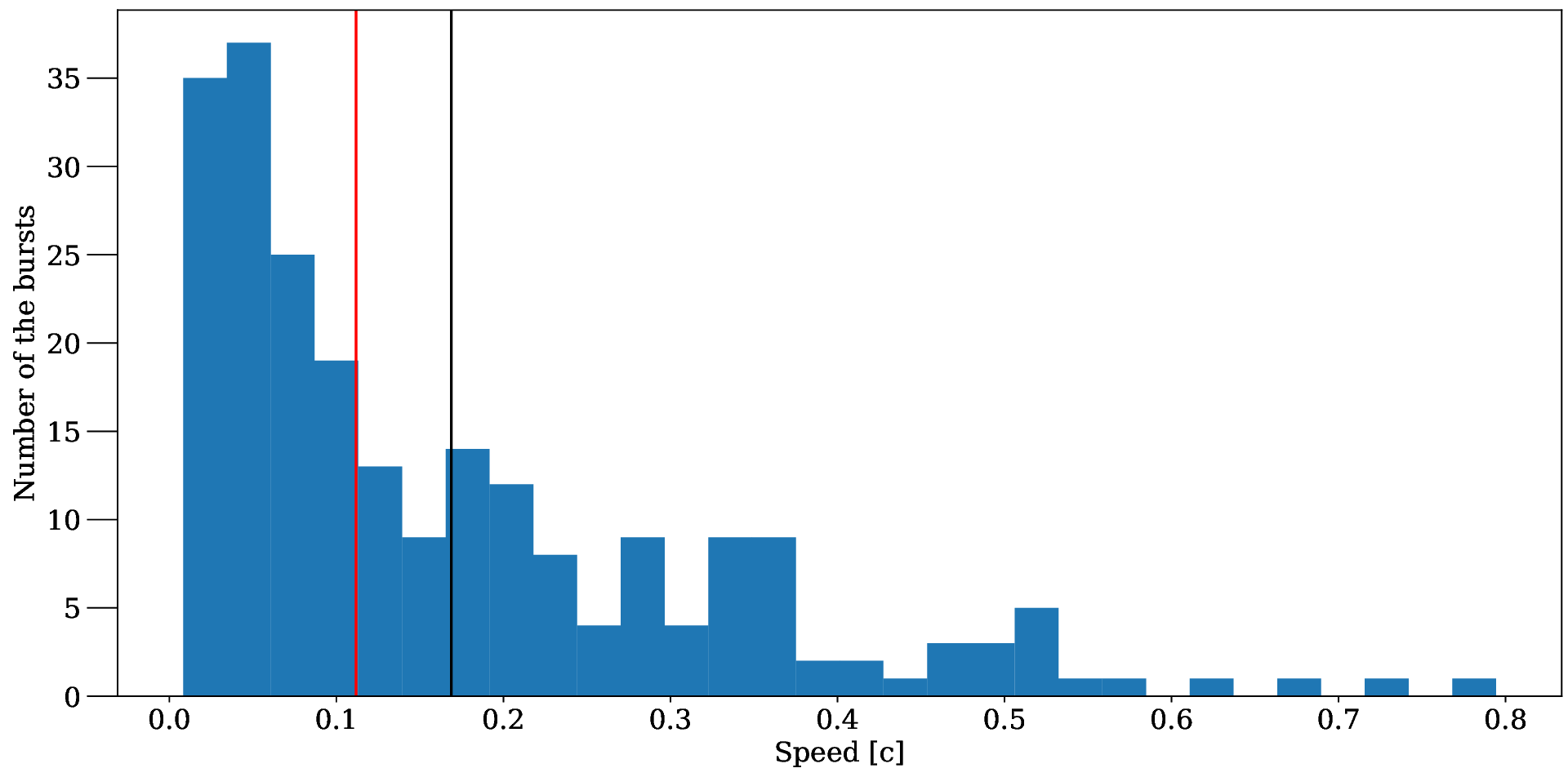}
    \caption{{Estimated electron beam speeds (relative to the speed of light, $c$) for type III bursts, using Eq.~(\ref{for:for4}) and the Newkirk density model. The red line is the median speed value, and the black line is the mean speed value. The harmonic index $s =  1$ (top) and $s =  2$ (bottom), see Eq.~(\ref{for:for2}).}}
    \label{fig:vel}
\end{figure}

{Thus, for Type IIIs, we estimated electron beam speeds ranging from 0.005 to 0.477 of the speed of light (mean: 0.102,  median: 0.06)  for the fundamental frequency ($s=1$), and 0.008 to 0.794 of the speed of light  (mean: 0.168, median: 0.111) for its second harmonic ($s=2$), as shown in Fig.~\ref{fig:vel}.
}  
 
\section{Conclusions} \label{Sec:con}

In this paper, we present a catalogue of solar radio bursts detected with the LV614 LOFAR station in the stand-alone mode from June 2022 to August 2025 in the 29--60~MHz frequency range. 
In total, the catalogue includes 335 solar radio bursts {\citep{steinbergs_2026_20157853}}. Radio bursts of all five main types, I--V, were detected. Specifically, the catalogue contains parameters of  23, 8, 293, 6 and 5 of Type I, II, III, IV, and V, respectively.
The detected burst peak flux ranges from 0.3~SFU to 234~SFU.
The catalogue provides information about the peak flux and frequency, and the height of the source above the solar surface, estimated under the assumption that the emission frequency coincides with the local electron plasma frequency and with the use of the Newkirk density stratification model \citep{2021SoPh..296...57K}. For Type II and III bursts, we also give the estimates of the {instantaneous} frequency drifts.  
The average hourly detection rates for different types of bursts, 0.02, 0.007, 0.264, 0.005, 0.005 for Type I, II, III, IV, and V, respectively, are consistent with the previously reported.

The numbers of Type I and III events are found to decrease with the peak flux. Approximating this dependence with a power-law function, the values of the power-law indices, $-0.9$ and $-1.25$, are found to be consistent with previous statistical studies, which typically report power-law slopes between around unity to $\sim2$, depending on observing frequency, instrument sensitivity, and event definition \citep[see e.g.][and references therein]{suresh2017wavelet, reid2014review}.
{As pointed out in Sec.~\ref{s:peak_flux_dens}, the derived power-law indices may be influenced by the frequency-dependent sensitivity of LOFAR and the detection threshold. To mitigate this in future work, the power-law indices can be obtained with a maximum-likelihood estimator for a truncated power law with an explicit lower bound $x_{\min}$ defined as the minimum flux value above which power-law behaviour is observed}  \citep{Clauset2009}.
{The value of $x_{\min}$ can be selected through the Kolmogorov–Smirnov statistic minimisation (to estimate how good the power law fits the data), with uncertainties estimated from $N$ bootstrap resamples of synthetic datasets. Fitting only above $x_{\min}$ will remove the incompleteness-affected part of the distribution by construction. One can also explore the stability of the index against the choice of the lower bound: for example, varying $x_{\min}$ between $S_c$ (limit for the flux above which the detection efficiency exceeds $90\%$ estimated from synthetic data) and $5S_c$ is expected to change the index by less than 0.1} \citep[see][]{ryan2016effects}{, which would confirm that the fitted slope is not driven by the truncation.}

The distances of the radio sources from the Sun, estimated from the peak frequency, range from 1.4 to about $2R_\odot$ from the solar centre. Type II--V bursts do not show a preferential formation height, whereas Type I bursts appear more likely to occur at lower heights.
{The Type III solar bursts show electron beam speeds ranging from 0.005 to 0.477 of the speed of light (mean: 0.102,  median: 0.06)  for the fundamental frequency ($s=1$), and 0.008 to 0.794 of the speed of light  (mean: 0.168, median: 0.111) for its second harmonic ($s=2$).
} 
{In general, excluding the assumed coronal density model, we envision LOFAR's frequency resolution and assumed emission harmonic as the main potential sources of uncertainty in our analysis. In particular, burst formation heights are computed from the detected peak frequencies. All bursts have the same frequency resolution of 195 kHz. Each 195-kHz subband has a specific centre frequency which determines the measured peak frequency that we use for estimating the formation heights. Thus, the estimated height difference between the start and end frequencies of each subband is 0.0086 $R/R_{\odot}$, resulting in less than 1\,{\%} error in the estimated heights caused by LOFAR's frequency resolution.
To account for the unknown emission harmonic number, in this work, we computed the formation heights and beam speed for different harmonic numbers $s=1$ and 2, see Figures \ref{fig:r} and \ref{fig:vel} which show the histograms of estimated heights and speeds for both fundamental and harmonic components of the type III solar bursts.}

The statistics of peak fluxes and peak frequencies have several important implications. Knowing the LOFAR sensitivity, we can estimate the probability of detecting different types of radio bursts on Sun-like stars using LOFAR single-station observations. This information is necessary for planning observational campaigns and estimating observing time allocations for the search for stellar flares and CMEs, particularly in the context of determining the habitable zones of exoplanets. Furthermore, the catalogue can be used as training data for machine-learning algorithms for real-time solar-burst detection and classification. In addition, in combination with other catalogues of extreme space-weather events, the catalogue provides a basis for various multi-wavelength and multi-instrument case studies.

In the future, we plan to expand the Irbene observatory capability to study the Sun by extending solar burst observations from 10 MHz to 11 GHz, combining LOFAR observations with RT~32 and e-CALLISTO.
This would increase the number of detected events and create a ground for performing a comprehensive analysis of solar radio bursts and revealing physical mechanisms responsible for this phenomenon \citep[e.g.,][]{bezrukovs}. Furthermore, increasing time and spectral resolution would be beneficial for the study of fine structures of solar radio bursts.

%

%

%
\appendix
\section{Solar bursts parameters}\label{appendix-A}
\scriptsize
\setlength{\LTpre}{0pt}
\setlength{\LTpost}{0pt}

\begin{longtable}{c c c c c c c c c}
\caption{Solar radio bursts detected with the LV614 LOFAR station from June 2022 to August 2025}
\label{tab:obs_full_list}\\

\toprule
ID & Date & Start time & Stop time & Peak frequency & Peak flux & Type & Source height & Drift rate \\
   &      &            &           & [MHz]          & [SFU]     &      & [$R/R_{\odot}$] & [MHz s$^{-1}$] \\
\midrule
\endfirsthead

\toprule
ID & Date & Start time & Stop time & Peak frequency & Peak flux & Type & Source height & Drift rate \\
   &      &            &           & [MHz]          & [SFU]     &      & [$R/R_{\odot}$] & [MHz s$^{-1}$] \\
\midrule
\endhead

\midrule
\multicolumn{9}{r}{Continued on next page}\\
\endfoot

\bottomrule
\endlastfoot

1 & 2022-08-13 & 12:45 & 12:55 & 41.02 & 35.239 & I & 1.602 & - \\
2 & 2022-09-16 & 10:18 & 10:35 & 59.18 & 17.598 & I & 1.433 & - \\
3 & 2022-09-16 & 09:41 & 10:16 & 53.52 & 15.003 & I & 1.476 & - \\
4 & 2022-09-16 & 10:36 & 11:08 & 60.55 & 41.288 & I & 1.424 & - \\
5 & 2023-05-09 & 07:51 & 08:10 & 52.15 & 11.462 & I & 1.487 & - \\
6 & 2023-05-09 & 08:32 & 08:35 & 46.09 & 3.219 & I & 1.544 & - \\
7 & 2023-05-09 & 08:36 & 08:56 & 47.27 & 8.605 & I & 1.532 & - \\
8 & 2023-05-09 & 09:30 & 09:33 & 53.91 & 3.309 & I & 1.473 & - \\
9 & 2023-05-09 & 09:37 & 09:42 & 48.63 & 6.150 & I & 1.519 & - \\
10 & 2023-05-09 & 09:43 & 09:49 & 60.74 & 10.091 & I & 1.422 & - \\
11 & 2023-05-10 & 06:10 & 06:55 & 40.04 & 3.338 & I & 1.615 & - \\
12 & 2023-05-10 & 04:40 & 05:29 & 44.53 & 2.378 & I & 1.561 & - \\
13 & 2023-05-10 & 05:25 & 05:45 & 48.63 & 0.784 & I & 1.519 & - \\
14 & 2023-07-04 & 11:55 & 12:20 & 58.98 & 0.734 & I & 1.434 & - \\
15 & 2024-06-19 & 06:35 & 06:55 & 60.55 & 4.595 & I & 1.424 & - \\
16 & 2024-08-07 & 14:09 & 14:13 & 50.78 & 7.008 & I & 1.499 & - \\
17 & 2024-10-03 & 08:34 & 09:53 & 34.57 & 118.518 & I & 1.696 & - \\
18 & 2024-10-03 & 10:50 & 12:10 & 29.49 & 16.825 & I & 1.793 & - \\
19 & 2024-11-01 & 10:20 & 10:37 & 40.23 & 137.453 & I & 1.612 & - \\
20 & 2025-02-21 & 12:20 & 12:32 & 29.49 & 6.543 & I & 1.793 & - \\
21 & 2025-06-17 & 07:43 & 07:46 & 35.94 & 36.183 & I & 1.674 & - \\
22 & 2025-06-17 & 07:51 & 09:29 & 37.11 & 20.902 & I & 1.656 & - \\
23 & 2025-07-25 & 09:53 & 11:14 & 36.52 & 53.554 & I & 1.665 & - \\

1 & 2024-07-03 & 07:42 & 08:04 & 36.72 & 132.183 & II & 1.662 & 0.127 \\
2 & 2024-07-16 & 13:25 & 13:55 & 40.62 & 191.889 & II & 1.607 & 0.017 \\
3 & 2024-07-17 & 06:39 & 06:45 & 52.34 & 4.366 & II & 1.486 & 0.048 \\
4 & 2024-07-17 & 07:00 & 07:06 & 60.74 & 5.296 & II & 1.422 & 0.034 \\
5 & 2024-07-23 & 14:10 & 14:44 & 32.03 & 134.059 & II & 1.741 & 0.021 \\
6 & 2024-07-24 & 07:45 & 08:02 & 41.41 & 82.044 & II & 1.597 & 0.035 \\
7 & 2024-08-07 & 13:52 & 14:09 & 30.27 & 216.088 & II & 1.642 & - \\
8 & 2024-08-14 & 06:58 & 07:18 & 45.31 & 18.228 & II & 1.552 & 0.019 \\

1 & 2023-01-04 & 09:25 & 09:25 & 30.47 & 17.941 & III & 1.772 & - \\
2 & 2023-01-04 & 11:25 & 11:31 & 44.53 & 63.783 & III & 1.561 & 7.780 \\
3 & 2023-01-04 & 13:11 & 13:11 & 53.32 & 0.350 & III & 1.477 & - \\
4 & 2022-06-15 & 07:26 & 07:28 & 47.46 & 18.904 & III & 1.530 & 8.050 \\
5 & 2022-08-13 & 12:44 & 12:45 & 60.74 & 38.510 & III & 1.422 & 1.803 \\
6 & 2022-11-09 & 10:50 & 10:51 & 49.61 & 10.696 & III & 1.510 & 10.435 \\
7 & 2022-11-09 & 11:51 & 11:52 & 31.25 & 3.250 & III & 1.756 & - \\
8 & 2022-11-22 & 12:26 & 12:34 & 45.51 & 88.487 & III & 1.550 & 3.795 \\
9 & 2022-11-22 & 12:48 & 12:57 & 37.11 & 5.302 & III & 1.656 & 5.971 \\
10 & 2022-11-15 & 10:16 & 10:16 & 29.30 & 1.000 & III & 1.797 & - \\
11 & 2022-11-15 & 10:58 & 11:07 & 29.88 & 3.925 & III & 1.784 & - \\
12 & 2022-11-15 & 13:57 & 13:57 & 30.08 & 1.926 & III & 1.780 & - \\
13 & 2022-10-26 & 12:37 & 12:46 & 48.63 & 31.576 & III & 1.519 & 2.186 \\
14 & 2022-12-07 & 08:46 & 08:47 & 39.84 & 56.030 & III & 1.617 & 1.378 \\
15 & 2024-06-11 & 08:40 & 08:41 & 21.68 & 14.693 & III & 2.017 & 23.730 \\
16 & 2024-06-11 & 16:38 & 16:39 & 62.70 & 58.388 & III & 1.410 & 6.934 \\
17 & 2024-06-11 & 17:23 & 17:23 & 50.39 & 3.862 & III & 1.503 & - \\
18 & 2024-06-26 & 02:18 & 02:21 & 37.11 & 54.685 & III & 1.656 & 2.246 \\
19 & 2024-06-26 & 10:16 & 10:27 & 50.59 & 46.277 & III & 1.501 & 2.539 \\
20 & 2024-06-25 & 08:20 & 08:23 & 60.74 & 27.160 & III & 1.422 & 5.241 \\
21 & 2024-06-25 & 09:26 & 09:29 & 35.74 & 44.361 & III & 1.677 & 2.644 \\
22 & 2024-06-25 & 09:47 & 09:56 & 36.13 & 118.783 & III & 1.671 & 2.507 \\
23 & 2024-06-25 & 11:06 & 11:14 & 41.21 & 56.832 & III & 1.600 & 1.964 \\
24 & 2024-06-25 & 13:31 & 13:32 & 40.43 & 12.206 & III & 1.610 & 2.445 \\
25 & 2024-06-25 & 14:21 & 14:22 & 37.11 & 32.523 & III & 1.656 & 0.529 \\
26 & 2024-06-25 & 15:02 & 15:03 & 29.30 & 3.185 & III & 1.797 & 4.520 \\
27 & 2024-06-25 & 17:57 & 17:58 & 29.49 & 36.787 & III & 1.793 & 2.832 \\
28 & 2024-07-02 & 12:37 & 12:42 & 58.98 & 4.701 & III & 1.434 & 5.036 \\
29 & 2024-07-02 & 16:50 & 16:51 & 56.84 & 0.851 & III & 1.450 & 6.362 \\
30 & 2023-01-18 & 12:01 & 12:01 & 29.30 & 3.339 & III & 1.797 & 1.245 \\
31 & 2023-01-25 & 08:04 & 08:04 & 31.45 & 1.346 & III & 1.752 & - \\
32 & 2023-01-25 & 08:18 & 08:23 & 35.16 & 21.816 & III & 1.686 & 6.217 \\
33 & 2023-01-25 & 12:20 & 12:21 & 47.66 & 28.254 & III & 1.528 & 3.458 \\
34 & 2023-02-01 & 08:03 & 08:03 & 56.25 & 6.768 & III & 1.454 & 4.310 \\
35 & 2023-02-07 & 14:10 & 14:14 & 35.16 & 1.232 & III & 1.686 & - \\
36 & 2023-02-07 & 14:58 & 15:01 & 29.30 & 6.692 & III & 1.797 & - \\
37 & 2023-02-08 & 11:00 & 11:01 & 60.55 & 1.727 & III & 1.424 & - \\
38 & 2023-02-15 & 14:25 & 14:25 & 51.56 & 9.593 & III & 1.492 & 6.706 \\
39 & 2023-02-21 & 11:31 & 11:32 & 40.82 & 86.565 & III & 1.605 & 8.561 \\
40 & 2023-02-21 & 14:34 & 14:39 & 45.12 & 75.782 & III & 1.555 & 7.747 \\
41 & 2023-02-28 & 10:45 & 10:46 & 29.30 & 20.704 & III & 1.797 & 13.574 \\
42 & 2023-02-28 & 14:50 & 14:56 & 47.27 & 8.542 & III & 1.532 & 10.268 \\
43 & 2023-03-01 & 11:43 & 11:44 & 39.84 & 46.181 & III & 1.617 & 5.249 \\
44 & 2023-03-01 & 13:15 & 13:16 & 60.55 & 2.745 & III & 1.424 & 1.654 \\
45 & 2023-03-01 & 13:45 & 13:48 & 30.66 & 35.937 & III & 1.768 & 10.254 \\
46 & 2023-03-08 & 12:21 & 12:26 & 30.08 & 32.099 & III & 1.780 & 10.352 \\
47 & 2023-03-28 & 13:20 & 13:20 & 29.88 & 2.420 & III & 1.784 & 1.162 \\
48 & 2023-03-29 & 10:45 & 10:48 & 51.95 & 14.498 & III & 1.489 & 8.984 \\
49 & 2023-03-29 & 11:52 & 11:52 & 46.09 & 4.331 & III & 1.544 & 10.184 \\
50 & 2023-03-29 & 13:01 & 13:02 & 31.05 & 1.000 & III & 1.760 & - \\
51 & 2023-03-29 & 13:36 & 13:38 & 29.49 & 25.853 & III & 1.793 & - \\
52 & 2023-03-29 & 13:52 & 13:55 & 51.95 & 22.586 & III & 1.489 & 3.427 \\
53 & 2023-04-04 & 07:43 & 07:43 & 30.08 & 19.430 & III & 1.780 & 1.649 \\
54 & 2023-04-04 & 11:38 & 11:38 & 42.19 & 9.250 & III & 1.588 & 2.102 \\
55 & 2023-04-05 & 09:58 & 09:58 & 34.96 & 5.470 & III & 1.689 & 2.232 \\
56 & 2023-04-05 & 10:35 & 10:35 & 40.43 & 3.849 & III & 1.610 & 1.600 \\
57 & 2023-04-28 & 08:59 & 09:11 & 34.57 & 60.330 & III & 1.696 & 0.467 \\
58 & 2023-04-28 & 11:00 & 11:03 & 34.77 & 31.337 & III & 1.692 & 6.380 \\
59 & 2023-04-28 & 11:31 & 11:38 & 33.40 & 12.410 & III & 1.716 & 7.064 \\
60 & 2023-05-09 & 07:39 & 07:42 & 31.25 & 67.951 & III & 1.756 & 5.134 \\
61 & 2023-05-09 & 10:48 & 10:50 & 36.33 & 58.798 & III & 1.668 & 1.207 \\
62 & 2023-05-09 & 12:02 & 12:05 & 39.26 & 94.139 & III & 1.625 & 7.357 \\
63 & 2023-05-09 & 14:01 & 14:02 & 30.27 & 136.375 & III & 1.776 & 1.158 \\
64 & 2023-05-09 & 17:45 & 17:46 & 54.69 & 2.194 & III & 1.466 & 14.209 \\
65 & 2023-05-09 & 18:28 & 18:37 & 30.27 & 36.463 & III & 1.776 & 0.652 \\
66 & 2023-05-10 & 03:10 & 03:14 & 37.11 & 73.966 & III & 1.656 & 1.052 \\
67 & 2023-05-10 & 04:08 & 04:16 & 49.80 & 34.054 & III & 1.508 & 5.539 \\
68 & 2023-05-10 & 04:08 & 04:16 & 49.80 & 34.054 & III & 1.508 & 5.539 \\
69 & 2023-05-10 & 05:53 & 05:55 & 60.74 & 19.261 & III & 1.422 & 7.096 \\
70 & 2023-05-10 & 06:58 & 06:59 & 29.30 & 4.357 & III & 1.797 & - \\
71 & 2023-05-10 & 07:22 & 07:23 & 46.29 & 45.997 & III & 1.542 & 1.764 \\
72 & 2023-05-10 & 08:07 & 08:08 & 44.92 & 69.279 & III & 1.557 & 2.597 \\
73 & 2023-05-10 & 09:33 & 09:33 & 50.20 & 23.506 & III & 1.504 & 6.343 \\
74 & 2023-05-23 & 08:54 & 08:56 & 44.53 & 22.790 & III & 1.561 & 0.243 \\
75 & 2023-05-23 & 11:18 & 11:19 & 36.91 & 83.527 & III & 1.659 & 9.528 \\
76 & 2023-05-23 & 11:57 & 11:58 & 43.95 & 45.170 & III & 1.567 & 0.391 \\
77 & 2023-05-24 & 01:55 & 01:58 & 49.22 & 0.599 & III & 1.513 & - \\
78 & 2023-05-24 & 02:07 & 02:08 & 52.54 & 0.916 & III & 1.484 & - \\
79 & 2023-05-24 & 03:57 & 03:57 & 40.23 & 21.273 & III & 1.612 & 5.246 \\
80 & 2023-05-24 & 04:38 & 04:43 & 47.85 & 16.201 & III & 1.527 & 5.859 \\
81 & 2023-05-24 & 12:54 & 12:55 & 30.27 & 137.814 & III & 1.776 & 0.367 \\
82 & 2023-05-30 & 12:50 & 12:50 & 42.38 & 77.555 & III & 1.586 & 3.145 \\
83 & 2023-05-30 & 13:38 & 13:44 & 42.38 & 0.671 & III & 1.586 & - \\
84 & 2023-05-30 & 14:40 & 14:40 & 56.25 & 9.617 & III & 1.454 & 2.916 \\
85 & 2023-05-30 & 18:48 & 19:01 & 29.88 & 0.616 & III & 1.784 & - \\
86 & 2023-06-27 & 08:20 & 08:21 & 50.00 & 16.715 & III & 1.506 & 4.529 \\
87 & 2023-06-27 & 14:11 & 14:12 & 50.39 & 1.471 & III & 1.503 & - \\
88 & 2023-06-28 & 03:18 & 03:18 & 38.28 & 51.018 & III & 1.639 & 3.190 \\
89 & 2023-06-28 & 04:21 & 04:25 & 45.51 & 22.420 & III & 1.550 & 1.488 \\
90 & 2023-06-28 & 05:55 & 06:05 & 36.52 & 119.172 & III & 1.665 & 0.884 \\
91 & 2023-06-28 & 06:02 & 06:03 & 60.74 & 0.387 & III & 1.422 & - \\
92 & 2023-07-04 & 07:32 & 07:32 & 53.12 & 1.939 & III & 1.479 & 5.469 \\
93 & 2023-07-04 & 13:57 & 13:58 & 29.88 & 8.219 & III & 1.784 & 9.208 \\
94 & 2023-07-04 & 14:26 & 14:26 & 51.95 & 4.276 & III & 1.489 & 4.688 \\
95 & 2023-07-05 & 08:09 & 08:09 & 47.66 & 55.832 & III & 1.528 & 3.797 \\
96 & 2023-07-05 & 09:04 & 09:14 & 32.42 & 7.127 & III & 1.734 & 2.020 \\
97 & 2023-07-05 & 10:12 & 10:18 & 37.30 & 80.647 & III & 1.653 & 3.027 \\
98 & 2023-07-05 & 11:27 & 11:28 & 50.39 & 12.529 & III & 1.503 & 4.427 \\
99 & 2023-07-05 & 12:24 & 12:26 & 35.94 & 174.314 & III & 1.674 & 1.777 \\
100 & 2023-07-05 & 13:47 & 13:55 & 60.35 & 10.840 & III & 1.425 & 4.102 \\
101 & 2023-07-11 & 09:32 & 09:32 & 36.33 & 121.140 & III & 1.668 & 1.786 \\
102 & 2023-07-11 & 11:17 & 11:20 & 32.42 & 234.063 & III & 1.734 & 0.797 \\
103 & 2023-08-08 & 09:27 & 09:29 & 29.88 & 185.546 & III & 1.784 & 0.433 \\
104 & 2023-08-08 & 10:52 & 10:53 & 34.77 & 6.572 & III & 1.692 & 9.375 \\
105 & 2023-08-08 & 11:40 & 11:41 & 31.05 & 9.325 & III & 1.760 & 1.426 \\
106 & 2023-08-08 & 09:27 & 09:29 & 29.88 & 185.546 & III & 1.784 & 0.433 \\
107 & 2023-08-08 & 10:52 & 10:53 & 34.77 & 6.572 & III & 1.692 & 9.375 \\
108 & 2023-08-08 & 11:40 & 11:41 & 31.05 & 9.325 & III & 1.760 & 1.426 \\
109 & 2023-10-16 & 11:40 & 12:27 & 33.59 & 45.710 & III & 1.712 & 5.957 \\
110 & 2023-10-17 & 08:45 & 08:46 & 37.89 & 53.845 & III & 1.644 & 6.315 \\
111 & 2023-09-05 & 08:10 & 08:11 & 31.25 & 2.136 & III & 1.756 & 2.865 \\
112 & 2023-09-05 & 08:52 & 08:55 & 60.74 & 42.439 & III & 1.422 & 0.563 \\
113 & 2023-09-05 & 13:16 & 13:21 & 50.20 & 27.942 & III & 1.504 & 6.376 \\
114 & 2023-09-06 & 10:58 & 10:59 & 60.55 & 0.420 & III & 1.424 & - \\
115 & 2023-09-06 & 11:47 & 11:48 & 60.74 & 0.403 & III & 1.422 & - \\
116 & 2023-09-06 & 12:39 & 12:40 & 60.74 & 0.792 & III & 1.422 & - \\
117 & 2023-09-12 & 08:50 & 08:51 & 60.74 & 3.970 & III & 1.422 & 5.246 \\
118 & 2023-09-12 & 10:46 & 10:46 & 48.05 & 4.449 & III & 1.525 & 4.102 \\
119 & 2023-09-13 & 09:25 & 09:26 & 50.78 & 6.604 & III & 1.499 & 1.399 \\
120 & 2023-09-13 & 10:57 & 10:57 & 29.30 & 5.463 & III & 1.797 & 1.343 \\
121 & 2023-09-13 & 11:14 & 11:15 & 39.06 & 28.684 & III & 1.628 & 13.770 \\
122 & 2023-09-13 & 11:14 & 11:16 & 44.92 & 70.365 & III & 1.557 & 1.526 \\
123 & 2023-11-01 & 08:47 & 08:48 & 33.20 & 159.641 & III & 1.719 & 9.212 \\
124 & 2023-11-01 & 09:35 & 09:37 & 29.30 & 38.964 & III & 1.797 & - \\
125 & 2023-11-01 & 09:55 & 09:56 & 29.30 & 27.023 & III & 1.797 & - \\
126 & 2023-11-08 & 06:05 & 06:05 & 29.88 & 17.238 & III & 1.784 & - \\
127 & 2023-11-08 & 07:37 & 07:37 & 54.10 & 4.065 & III & 1.471 & 9.459 \\
128 & 2023-11-08 & 11:09 & 11:10 & 29.69 & 49.907 & III & 1.789 & - \\
129 & 2023-11-08 & 11:44 & 11:52 & 34.96 & 27.693 & III & 1.689 & 7.878 \\
130 & 2023-11-14 & 12:51 & 12:51 & 52.15 & 16.424 & III & 1.487 & 3.996 \\
131 & 2023-11-14 & 13:53 & 13:54 & 33.40 & 4.708 & III & 1.716 & - \\
132 & 2023-11-15 & 12:10 & 12:11 & 55.27 & 4.238 & III & 1.462 & 10.124 \\
133 & 2023-11-28 & 08:34 & 08:35 & 34.57 & 114.497 & III & 1.696 & - \\
134 & 2023-11-28 & 11:54 & 11:55 & 29.30 & 4.489 & III & 1.797 & - \\
135 & 2023-12-05 & 10:08 & 10:09 & 32.23 & 12.227 & III & 1.737 & - \\
136 & 2022-11-30 & 11:04 & 11:12 & 31.64 & 19.917 & III & 1.749 & 0.486 \\
137 & 2024-07-09 & 08:32 & 08:32 & 60.35 & 0.619 & III & 1.425 & - \\
138 & 2024-07-09 & 13:31 & 13:31 & 48.24 & 8.672 & III & 1.523 & 1.810 \\
139 & 2024-07-09 & 15:33 & 15:34 & 54.49 & 2.208 & III & 1.468 & - \\
140 & 2024-06-05 & 09:14 & 09:16 & 25.20 & 141.883 & III & 1.901 & 14.328 \\
141 & 2024-06-18 & 12:12 & 12:18 & 49.02 & 13.274 & III & 1.515 & 1.348 \\
142 & 2024-06-18 & 13:11 & 13:12 & 35.16 & 1.806 & III & 1.686 & 5.552 \\
143 & 2024-06-18 & 13:36 & 13:38 & 55.08 & 2.721 & III & 1.463 & 3.581 \\
144 & 2024-06-18 & 14:52 & 14:53 & 39.65 & 2.229 & III & 1.620 & 7.096 \\
145 & 2024-06-18 & 15:35 & 15:37 & 54.69 & 4.921 & III & 1.466 & 5.887 \\
146 & 2024-06-18 & 17:53 & 17:57 & 53.52 & 0.777 & III & 1.476 & - \\
147 & 2024-06-19 & 05:02 & 05:03 & 60.55 & 4.288 & III & 1.424 & 0.478 \\
148 & 2024-06-19 & 05:56 & 05:57 & 39.84 & 31.194 & III & 1.617 & 3.246 \\
149 & 2024-06-19 & 07:34 & 07:36 & 39.26 & 55.444 & III & 1.625 & 2.630 \\
150 & 2024-06-19 & 09:04 & 09:05 & 29.30 & 35.002 & III & 1.797 & 1.857 \\
151 & 2024-07-03 & 07:35 & 07:38 & 41.21 & 36.463 & III & 1.600 & 0.439 \\
152 & 2024-07-03 & 10:54 & 10:56 & 44.14 & 64.183 & III & 1.565 & 7.682 \\
153 & 2024-07-16 & 09:19 & 09:20 & 60.74 & 2.014 & III & 1.422 & 5.241 \\
154 & 2024-07-16 & 13:22 & 13:23 & 60.35 & 0.407 & III & 1.425 & - \\
155 & 2024-07-17 & 02:37 & 02:38 & 49.22 & 1.066 & III & 1.513 & - \\
156 & 2024-07-17 & 03:55 & 03:56 & 51.76 & 2.486 & III & 1.491 & 6.138 \\
157 & 2024-07-17 & 04:51 & 04:51 & 60.74 & 3.128 & III & 1.422 & 5.748 \\
158 & 2024-07-17 & 06:09 & 06:12 & 42.38 & 120.525 & III & 1.586 & 0.573 \\
159 & 2024-07-17 & 06:25 & 06:27 & 59.77 & 1.907 & III & 1.429 & - \\
160 & 2024-07-23 & 08:06 & 08:09 & 45.12 & 8.427 & III & 1.555 & 0.840 \\
161 & 2024-07-23 & 14:03 & 14:08 & 33.01 & 108.718 & III & 1.723 & 1.061 \\
162 & 2024-07-24 & 04:56 & 04:56 & 60.74 & 19.075 & III & 1.422 & 1.583 \\
163 & 2024-07-24 & 05:55 & 05:56 & 60.74 & 10.849 & III & 1.422 & 5.748 \\
164 & 2024-07-24 & 06:54 & 06:59 & 40.23 & 67.043 & III & 1.612 & 1.900 \\
165 & 2024-07-24 & 07:32 & 07:38 & 31.45 & 172.993 & III & 1.752 & 0.259 \\
166 & 2024-07-24 & 10:46 & 10:53 & 30.08 & 178.618 & III & 1.780 & 0.326 \\
167 & 2024-07-24 & 11:17 & 11:20 & 45.70 & 46.108 & III & 1.548 & 4.492 \\
168 & 2024-07-24 & 11:51 & 11:56 & 42.58 & 81.795 & III & 1.583 & 10.282 \\
169 & 2024-07-24 & 12:49 & 12:59 & 29.49 & 44.668 & III & 1.793 & 2.995 \\
170 & 2024-08-06 & 08:43 & 08:46 & 32.62 & 176.717 & III & 1.730 & 6.413 \\
171 & 2024-08-06 & 09:29 & 09:30 & 39.06 & 112.139 & III & 1.628 & 2.693 \\
172 & 2024-08-06 & 10:02 & 10:03 & 35.55 & 173.947 & III & 1.680 & 1.511 \\
173 & 2024-08-07 & 04:24 & 04:25 & 60.74 & 13.770 & III & 1.422 & 3.398 \\
174 & 2024-08-07 & 10:33 & 10:39 & 45.31 & 88.446 & III & 1.552 & 2.413 \\
175 & 2024-08-07 & 13:33 & 13:35 & 60.16 & 19.340 & III & 1.426 & 0.262 \\
176 & 2024-08-13 & 08:02 & 08:03 & 45.70 & 108.543 & III & 1.548 & 1.150 \\
177 & 2024-08-13 & 11:59 & 11:59 & 41.41 & 1.500 & III & 1.597 & 5.859 \\
178 & 2024-08-13 & 14:49 & 14:50 & 57.42 & 9.021 & III & 1.446 & 4.150 \\
179 & 2024-08-13 & 16:09 & 16:11 & 39.84 & 104.152 & III & 1.617 & 1.464 \\
180 & 2024-08-14 & 07:55 & 07:56 & 44.92 & 25.279 & III & 1.557 & 0.426 \\
181 & 2024-08-14 & 08:19 & 08:19 & 31.64 & 69.134 & III & 1.749 & 1.536 \\
182 & 2024-08-14 & 13:32 & 13:33 & 38.09 & 2.458 & III & 1.642 & 4.381 \\
183 & 2024-08-20 & 08:33 & 08:37 & 31.84 & 52.453 & III & 1.745 & 3.627 \\
184 & 2024-08-20 & 09:10 & 09:10 & 47.27 & 49.299 & III & 1.532 & 1.747 \\
185 & 2024-08-20 & 11:43 & 11:44 & 55.86 & 12.598 & III & 1.457 & 1.473 \\
186 & 2024-08-20 & 14:24 & 14:24 & 51.76 & 1.155 & III & 1.491 & 5.301 \\
187 & 2024-08-21 & 11:26 & 11:27 & 60.74 & 2.690 & III & 1.422 & 9.180 \\
188 & 2024-08-28 & 04:30 & 04:31 & 60.74 & 0.252 & III & 1.422 & - \\
189 & 2024-08-28 & 08:57 & 09:05 & 60.74 & 52.671 & III & 1.422 & 1.104 \\
190 & 2024-09-10 & 12:01 & 12:02 & 37.89 & 4.290 & III & 1.644 & 6.673 \\
191 & 2024-09-10 & 12:01 & 12:01 & 37.89 & 4.290 & III & 1.644 & 6.673 \\
192 & 2024-09-11 & 09:23 & 09:23 & 29.30 & 6.694 & III & 1.797 & - \\
193 & 2024-09-11 & 13:36 & 13:37 & 43.95 & 42.244 & III & 1.567 & 2.006 \\
194 & 2024-09-16 & 07:23 & 07:23 & 60.16 & 5.858 & III & 1.426 & 1.197 \\
195 & 2024-09-23 & 07:54 & 07:56 & 54.88 & 9.340 & III & 1.465 & 5.241 \\
196 & 2024-09-23 & 09:27 & 09:27 & 29.49 & 11.854 & III & 1.793 & 1.844 \\
197 & 2024-09-24 & 11:35 & 11:36 & 44.92 & 30.513 & III & 1.557 & 3.848 \\
198 & 2024-09-25 & 07:57 & 07:58 & 32.03 & 9.495 & III & 1.741 & 2.226 \\
199 & 2024-09-25 & 12:24 & 12:24 & 60.55 & 22.388 & III & 1.424 & 0.393 \\
200 & 2024-09-25 & 14:16 & 14:22 & 29.30 & 7.023 & III & 1.797 & 7.715 \\
201 & 2024-10-02 & 06:57 & 07:03 & 38.09 & 15.005 & III & 1.642 & - \\
202 & 2024-10-02 & 08:41 & 08:42 & 35.94 & 65.570 & III & 1.674 & 0.518 \\
203 & 2024-10-02 & 12:19 & 12:20 & 47.66 & 32.960 & III & 1.528 & 1.849 \\
204 & 2024-10-02 & 12:49 & 12:52 & 30.08 & 5.143 & III & 1.780 & - \\
205 & 2024-10-02 & 14:02 & 14:08 & 29.88 & 10.032 & III & 1.784 & - \\
206 & 2024-10-03 & 06:22 & 06:23 & 38.67 & 3.341 & III & 1.633 & - \\
207 & 2024-10-03 & 07:13 & 07:13 & 36.91 & 87.818 & III & 1.659 & 1.866 \\
208 & 2024-10-03 & 11:16 & 11:16 & 55.08 & 1.668 & III & 1.463 & 0.313 \\
209 & 2024-10-03 & 12:05 & 12:05 & 36.13 & 8.867 & III & 1.671 & 5.241 \\
210 & 2024-10-07 & 09:41 & 09:42 & 50.78 & 26.123 & III & 1.499 & 2.620 \\
211 & 2024-10-08 & 10:03 & 10:06 & 36.72 & 25.055 & III & 1.662 & 2.507 \\
212 & 2024-10-08 & 10:53 & 10:54 & 59.38 & 1.567 & III & 1.432 & - \\
213 & 2024-10-23 & 09:02 & 09:06 & 32.62 & 72.694 & III & 1.730 & 5.794 \\
214 & 2024-10-23 & 09:50 & 09:56 & 60.74 & 37.638 & III & 1.422 & 1.877 \\
215 & 2024-10-23 & 10:42 & 10:43 & 60.16 & 19.764 & III & 1.426 & 11.189 \\
216 & 2024-10-24 & 08:44 & 08:45 & 34.57 & 124.699 & III & 1.696 & 1.311 \\
217 & 2024-10-24 & 09:19 & 09:19 & 34.18 & 59.417 & III & 1.702 & 9.668 \\
218 & 2024-10-24 & 10:52 & 10:54 & 60.74 & 13.142 & III & 1.422 & 3.931 \\
219 & 2024-10-24 & 13:46 & 13:46 & 29.30 & 13.901 & III & 1.797 & - \\
220 & 2024-10-31 & 11:44 & 11:45 & 29.30 & 23.309 & III & 1.797 & 1.209 \\
221 & 2024-10-31 & 11:47 & 11:48 & 41.41 & 81.441 & III & 1.597 & 2.462 \\
222 & 2024-10-31 & 12:52 & 12:54 & 45.70 & 12.612 & III & 1.548 & 7.975 \\
223 & 2024-10-31 & 13:05 & 13:07 & 38.09 & 99.817 & III & 1.642 & 1.506 \\
224 & 2024-11-01 & 09:39 & 09:44 & 34.18 & 60.632 & III & 1.702 & 9.947 \\
225 & 2024-11-01 & 12:54 & 12:57 & 37.50 & 3.589 & III & 1.650 & - \\
226 & 2024-11-04 & 10:38 & 10:40 & 29.49 & 75.327 & III & 1.793 & - \\
227 & 2024-11-04 & 11:53 & 11:56 & 38.28 & 34.284 & III & 1.639 & 10.770 \\
228 & 2024-11-04 & 12:57 & 12:58 & 29.88 & 12.012 & III & 1.784 & 1.435 \\
229 & 2024-11-05 & 09:42 & 09:44 & 34.18 & 69.688 & III & 1.702 & 11.021 \\
230 & 2024-11-13 & 08:50 & 08:51 & 29.69 & 22.486 & III & 1.789 & 7.959 \\
231 & 2024-11-14 & 09:18 & 09:22 & 31.05 & 71.677 & III & 1.760 & - \\
232 & 2024-11-26 & 09:30 & 09:30 & 60.74 & 30.405 & III & 1.422 & 1.448 \\
233 & 2024-11-27 & 09:43 & 09:44 & 39.06 & 15.512 & III & 1.628 & 7.031 \\
234 & 2025-01-23 & 11:01 & 11:07 & 33.59 & 15.891 & III & 1.712 & - \\
235 & 2025-01-23 & 12:32 & 12:39 & 54.30 & 10.072 & III & 1.470 & 9.445 \\
236 & 2025-01-23 & 13:53 & 13:57 & 29.69 & 3.039 & III & 1.789 & - \\
237 & 2025-01-24 & 09:20 & 09:29 & 29.30 & 120.063 & III & 1.797 & 5.729 \\
238 & 2025-01-24 & 09:52 & 09:55 & 30.86 & 13.098 & III & 1.764 & 2.198 \\
239 & 2025-01-31 & 09:15 & 09:16 & 39.45 & 81.681 & III & 1.623 & 2.193 \\
240 & 2025-01-31 & 09:51 & 09:52 & 36.33 & 101.231 & III & 1.668 & 2.853 \\
241 & 2025-01-31 & 10:51 & 10:52 & 33.79 & 100.289 & III & 1.709 & 2.131 \\
242 & 2025-01-31 & 10:52 & 10:52 & 30.47 & 105.042 & III & 1.772 & 0.550 \\
243 & 2025-01-31 & 10:59 & 11:00 & 54.30 & 6.731 & III & 1.470 & 5.241 \\
244 & 2025-01-31 & 11:29 & 11:33 & 32.42 & 115.462 & III & 1.734 & 1.689 \\
245 & 2025-02-06 & 10:17 & 10:18 & 29.69 & 13.294 & III & 1.789 & - \\
246 & 2025-02-06 & 11:04 & 11:07 & 47.46 & 4.540 & III & 1.530 & 1.194 \\
247 & 2025-02-07 & 09:33 & 09:43 & 31.45 & 83.324 & III & 1.752 & 8.887 \\
248 & 2025-02-07 & 09:42 & 09:43 & 39.06 & 17.039 & III & 1.628 & 2.721 \\
249 & 2025-02-07 & 10:22 & 10:24 & 53.32 & 22.818 & III & 1.477 & 6.478 \\
250 & 2025-02-13 & 10:01 & 10:09 & 36.72 & 103.240 & III & 1.662 & 0.816 \\
251 & 2025-02-13 & 10:50 & 10:54 & 51.17 & 5.878 & III & 1.496 & 2.246 \\
252 & 2025-02-13 & 10:58 & 11:01 & 34.57 & 106.622 & III & 1.696 & 9.928 \\
253 & 2025-02-13 & 12:07 & 12:10 & 36.52 & 123.795 & III & 1.665 & 0.878 \\
254 & 2025-02-14 & 09:11 & 09:12 & 30.86 & 17.983 & III & 1.764 & 5.501 \\
255 & 2025-02-14 & 09:57 & 10:00 & 32.42 & 97.974 & III & 1.734 & 2.483 \\
256 & 2025-02-14 & 11:23 & 11:27 & 45.70 & 67.324 & III & 1.548 & 0.511 \\
257 & 2025-02-20 & 10:39 & 10:40 & 49.02 & 7.278 & III & 1.515 & 1.239 \\
258 & 2025-02-20 & 11:02 & 11:03 & 29.30 & 65.646 & III & 1.797 & - \\
259 & 2025-02-21 & 09:28 & 09:28 & 34.18 & 21.222 & III & 1.702 & 1.254 \\
260 & 2025-02-21 & 12:06 & 12:06 & 29.49 & 2.196 & III & 1.793 & - \\
261 & 2025-02-27 & 07:55 & 07:56 & 47.27 & 27.572 & III & 1.532 & 3.931 \\
262 & 2025-02-27 & 08:41 & 08:41 & 34.38 & 8.042 & III & 1.699 & 9.863 \\
263 & 2025-02-27 & 09:41 & 09:42 & 29.49 & 28.075 & III & 1.793 & - \\
264 & 2025-02-27 & 12:13 & 12:18 & 29.49 & 1.538 & III & 1.793 & - \\
265 & 2025-02-28 & 09:07 & 09:08 & 37.50 & 18.605 & III & 1.650 & - \\
266 & 2025-02-28 & 10:14 & 10:20 & 45.51 & 49.847 & III & 1.550 & 3.145 \\
267 & 2025-03-06 & 09:19 & 09:20 & 32.62 & 45.718 & III & 1.730 & 1.801 \\
268 & 2025-03-07 & 12:30 & 12:31 & 29.69 & 1.748 & III & 1.789 & - \\
269 & 2025-03-29 & 11:57 & 11:58 & 38.48 & 2.294 & III & 1.636 & 10.156 \\
270 & 2025-03-29 & 12:42 & 12:43 & 53.12 & 3.055 & III & 1.479 & 4.688 \\
271 & 2025-04-05 & 09:13 & 09:15 & 44.73 & 27.912 & III & 1.559 & 0.539 \\
272 & 2025-06-05 & 12:20 & 12:21 & 59.18 & 35.895 & III & 1.433 & 0.467 \\
273 & 2025-06-05 & 15:25 & 15:35 & 39.06 & 3.527 & III & 1.628 & 5.315 \\
274 & 2025-06-06 & 09:53 & 09:54 & 43.36 & 22.749 & III & 1.574 & 2.266 \\
275 & 2025-06-14 & 06:56 & 07:04 & 50.39 & 0.381 & III & 1.503 & - \\
276 & 2025-06-14 & 11:41 & 11:42 & 60.74 & 0.255 & III & 1.422 & - \\
277 & 2025-06-14 & 13:26 & 13:29 & 38.48 & 46.839 & III & 1.636 & 1.563 \\
278 & 2025-06-14 & 15:06 & 15:07 & 48.24 & 0.407 & III & 1.523 & - \\
279 & 2025-06-14 & 16:06 & 16:08 & 49.80 & 8.534 & III & 1.508 & 4.590 \\
280 & 2025-06-15 & 12:10 & 12:11 & 59.57 & 0.908 & III & 1.430 & - \\
281 & 2025-06-15 & 13:09 & 13:15 & 37.50 & 94.903 & III & 1.650 & 0.846 \\
282 & 2025-06-17 & 06:26 & 06:26 & 50.59 & 7.270 & III & 1.501 & 5.241 \\
283 & 2025-06-17 & 09:51 & 09:51 & 60.74 & 13.777 & III & 1.422 & 1.810 \\
284 & 2025-06-17 & 10:56 & 10:57 & 29.30 & 49.759 & III & 1.797 & 1.048 \\
285 & 2025-06-17 & 11:55 & 11:55 & 60.74 & 6.798 & III & 1.422 & 5.469 \\
286 & 2025-06-17 & 13:09 & 13:09 & 60.55 & 14.812 & III & 1.424 & 0.662 \\
287 & 2025-06-17 & 14:01 & 14:02 & 36.13 & 6.638 & III & 1.671 & - \\
288 & 2025-06-30 & 12:50 & 12:51 & 45.31 & 0.931 & III & 1.552 & - \\
289 & 2025-07-01 & 06:38 & 06:38 & 29.88 & 6.026 & III & 1.784 & 1.517 \\
290 & 2025-07-01 & 08:30 & 08:32 & 48.24 & 42.760 & III & 1.523 & 10.700 \\
291 & 2025-07-26 & 07:08 & 07:08 & 48.05 & 65.560 & III & 1.525 & 10.296 \\
292 & 2025-07-26 & 11:12 & 11:12 & 53.12 & 9.582 & III & 1.479 & 2.024 \\
293 & 2025-08-02 & 09:43 & 09:44 & 60.74 & 0.476 & III & 1.422 & - \\

1 & 2024-06-11 & 16:38 & 16:40 & 62.70 & 58.388 & IV & 1.410 & - \\
2 & 2023-05-23 & 11:20 & 17:57 & 37.50 & 80.463 & IV & 1.650 & - \\
3 & 2024-07-16 & 13:43 & 14:51 & 33.01 & 134.042 & IV & 1.723 & - \\
4 & 2024-07-17 & 07:22 & 07:46 & 60.74 & 12.780 & IV & 1.422 & - \\
5 & 2024-08-14 & 07:21 & 11:53 & 31.64 & 69.134 & IV & 1.749 & - \\
6 & 2024-11-01 & 10:34 & 12:25 & 40.23 & 149.176 & IV & 1.612 & - \\

1 & 2022-12-07 & 08:38 & 08:39 & 35.94 & 172.819 & V & 1.674 & - \\
2 & 2023-05-10 & 06:06 & 06:07 & 48.44 & 8.370 & V & 1.521 & - \\
3 & 2023-08-08 & 09:27 & 09:30 & 29.88 & 185.546 & V & 1.784 & - \\
4 & 2024-07-17 & 06:09 & 06:12 & 42.38 & 120.525 & V & 1.586 & - \\
5 & 2024-07-24 & 07:32 & 07:40 & 31.45 & 172.993 & V & 1.752 & - 

\end{longtable}

 \begin{acknowledgments}
 This research has been supported by the EU Recovery and Resilience Facility within Project No 5.2.1.1.i.0/2/24/I/CFLA/003 \lq\lq Implementation of consolidation and management changes at Riga Technical University, Liepaja University, Rezekne Academy of Technology, Latvian Maritime Academy and Liepaja Maritime College for the progress towards excellence in higher education, science and innovation\rq\rq\ academic career doctoral grant (ID 1001).
DK, VMN, and VB contributed to this work within the project \lq\lq Novel proxies and approaches for solar flare forecasting\rq\rq, Nr. lzp-2024/1-0023 (FLARE).
DK also thanks the UKRI Stephen Hawking Fellowship EP/Z535473/1.
LOFAR, the Low-Frequency Array designed and constructed by ASTRON, has facilities in several countries, which are owned by various parties (each with their own funding
sources), and that are collectively operated by the International LOFAR Telescope (ILT) foundation under a joint scientific policy. LOFAR station LV614 is hosted
and operated by the Ventspils Radio Astronomy Center, Ventspils University of Applied Sciences.
{We extend our thanks to the referee for constructive comments on improving this paper.}
 \end{acknowledgments}

%
%
\bibliography{main}{}
\bibliographystyle{aasjournal}

%
%
%
%

\end{document}